\documentclass[pra,superscriptaddress,preprint,byrevtex,showpacs,amsmath,amssymb]{revtex4}

\usepackage{graphicx}

\begin{document}

\title{A Model for an Irreversible Bias Current in the Superconducting
Qubit Measurement Process.}
\author{G.~D.~Hutchinson}
\email{gdh24@cam.ac.uk}
\affiliation{Hitachi Cambridge Laboratory, Hitachi Europe Ltd.,
Cambridge CB3 OHE, UK}
\author{C.~A.~Holmes}
\affiliation{Department of Mathematics, University of Queensland,
St.~Lucia, Queensland 4072, Australia}
\author{T.~M.~Stace}
\affiliation{Department of Applied Mathematics and Theoretical Physics, University
of Cambridge, Cambridge CB3 0WA, UK}
\author{T.~P.~Spiller}
\affiliation{Quantum Information Processing Group, Hewlett-Packard
Laboratories, Filton Road, Stoke Gifford, Bristol BS34 8QZ, UK}
\author{G.~J.~Milburn}
\affiliation{Centre for Quantum Computer Technology, Department
of Physics, University of Queensland, St.~Lucia, Queensland 4072,
Australia}
\author{S.~D.~Barrett\footnote{Current Address: Blackett Laboratory, Imperial College London, Prince Consort Road, London, SW7 2BW}}
\affiliation{Quantum Information Processing Group, Hewlett-Packard
Laboratories, Filton Road, Stoke Gifford, Bristol BS34 8QZ, UK}
\author{D.~G.~Hasko}
\affiliation{Microelectronics Research Centre, Cavendish Laboratory,
University of Cambridge, Cambridge CB3 0HE, UK}
\author{D.~A.~Williams}
\affiliation{Hitachi Cambridge Laboratory, Hitachi Europe Ltd.,
Cambridge CB3 OHE, UK}
\date{June 6, 2006}
\begin{abstract}
The superconducting charge-phase `Quantronium' qubit is considered
in order to develop a model for the measurement process used in
the experiment of Vion \textit{et.~al.} [Science \textbf{296} 886
(2002)]. For this model we propose a method for including the bias
current in the read-out process in a fundamentally irreversible
way, which to first order, is approximated by the Josephson junction
tilted-washboard potential phenomenology. The decohering bias current
is introduced in the form of a Lindblad operator and the Wigner
function for the current biased read-out Josephson junction is derived
and analyzed. During the read-out current pulse used in the Quantronium
experiment we find that the coherence of the qubit initially prepared
in a symmetric superposition state is lost at a time of 0.2 nanoseconds
after the bias current pulse has been applied. A timescale which is 
much shorter than the experimental readout time. Additionally we look
at the effect of Johnson-Nyquist noise with zero mean from the current source
during the qubit manipulation and show that the decoherence due
to the irreversible bias current description is an order of magnitude
smaller than that found through adding noise to the reversible tilted
washboard potential model. Our irreversible bias current model is
also applicable to the persistent current based qubits where the
state is measured according to its flux via a small inductance direct
current superconducting quantum interference device (DC-SQUID). 
\end{abstract}
\maketitle

\section{Introduction}

Quantum computers and the quantum algorithms that run on them have
been proposed as a technology to perform computational tasks not
tractable with classical computer circuits\cite{neilsen}. Recent
experiments have provided significant advances towards developing
the fundamental element of this technology, the quantum bit or qubit.
So far, qubit systems based on nuclear magnetic resonance\cite{{NMR1,
NMR2}} and ion traps\cite{{ionTraps, ionTraps2}} have been used
to show multiple qubit operation, whilst efficient linear optic
quantum computing\cite{KLM} has been demonstrated with the successful
operation of the two qubit controlled-not gate\cite{UQCNOT}. Experimental
advances have also been made in solid state systems which utilise
a wide variety of quantum effects in many different materials. The
main attraction of solid state systems is the possibility to
scale such technology using modern-day device fabrication techniques
once the implementation of component gates has been demonstrated.
Promising solid state systems include the use of phosphor dopants in silicon\cite{kane}, 
charge based quantum dots\cite{{blick, Fujisawa, Kouwenhouven}}, optically 
controlled exciton systems\cite{XLiScience301-809} as well as a variety of
systems based on the coherent electron state in superconducting
materials\cite{review}.

In these superconducting systems the implementation of single qubit
operation\cite{{nakamura, Martinis, Yu, delft, Vion}}, some with
single shot readout, has been demonstrated. Also devices with a
non-switchable inter-qubit interaction between two qubits have been
shown\cite{{paskin, Maryland}}, providing the initial evidence for
a two-qubit entangled state in these structures. To ensure scalability
to more complex configurations into the future there is a need to
identify ways to develop more accurate gates, provide higher fidelity
readout and ensure longer coherence times in the devices being developed.
For instance the `Quantronium' charge-phase qubit developed by Vion
\textit{et al.}\cite{Vion} was designed to be insensitive to first
order fluctuations in the external control parameters of the system
provided that the control parameters for the device, in this case
the voltage and applied flux, were used about an `optimal point'
of the system with this property. In this experiment the quality
factor of quantum coherence Q for the device, defined as the number
of elementary gate operations that could be performed before the
device state decoheres, was found to be of the order $10^{4}$.

In this paper we examine the readout process in the experiment of
Vion \textit{et.~al.} through the Lindblad operator formalism\cite{limblad}
and we introduce the bias current into the model in a fundamentally
irreversible way that acts to decohere the state of the qubit. Using this method we implement a heuristic model for the measurement process that is induced by the application of the bias current to the quantronium circuit. This model allows for the bias current to `count' the number of electrons that pass through the system during the measurement process and in doing so destroys the coherence between the different states of the system. Therefore to examine this model we are ignoring the typical terms which appear
in the system master equation that describe the widely known forms
of decoherence for the qubit through its coupling to the environment,
such as the ohmic dissipation of the leads\cite{leggett}. Our aim
is to gain further insight
into the role of the irreversible readout process and the decohering
process associated with its operation. 

The irreversible dynamics arising 
from the current bias provides a decoherence mechanism
that collapses the quantum superposition to a probabilistic mixture 
on a time scale shorter than the time for the state to tunnel out of the 
metastable qubit states into unbound states of the washboard
potential and create a voltage on the read-out voltmeter.
This means that the measurement of the system is performed during
the application of the bias current before any classical information
about the qubit state is returned to the experimentalist. Such 
`measurement induced decoherence' is analogous to that discussed 
in semiconducting systems\cite{stace:136802}. In addition
to this we also analyse the implications this irreversible current
source has for the effect of Johnson-Nyquist noise from the current source
during the qubit manipulation when the bias current has a zero mean
and intended to be decoupled form the device.

\section{The Current Biased Josephson Junction}

The measurement process in superconducting qubit structures such
as the Quantronium and the direct current superconducting quantum
interference device (DC-SQUID) (which is used to measure the persistent
current qubits\cite{delft} and proposed to measure magnetic nanoparticles\cite{spiller2})
rely upon the transition of a Josephson junction based system 
from the superconducting state into the voltage state, where the information 
associated with the effective critical current of the device provides 
the quantum state measurement. The semi-classical model for a single
Josephson junction is the one-dimensional analogy to a particle of mass $(\hbar /2e)^{2}C$ 
moving along the $\gamma $ axis in the potential\cite{tinkham} 
\begin{equation}
U( \gamma ) =E_{J}( 1-\cos \gamma ) -E_{b}\gamma ,
\label{XRef-Equation-9613151}
\end{equation}
where $E_{b}=I_{bias}\Phi _{0}/2\pi $. The $E_{b}\gamma
$ term describes the slope of the washboard potential, which has been used widely in the quantum regime\cite{leggett}. For instance,
it has been used to describe the escape rates of macroscopic tunnelling
events in current biased Josephson junctions\cite{{Martinis2, Martinis3}}.
The inclusion of the linear potential in Eq. (\ref{XRef-Equation-9613151}) to 
create the tilted washboard potential does not contribute any dephasing
term to the dynamics and implies that the measurement process is
intrinsically reversible. That is, by turning the current source on and 
then off again, the qubit is back in its initial state (provided a 
macroscopic quantum tunnelling event has not occurred).

In this paper we propose an alternate description of the current
bias in the Quantronium and other current biased systems such as
the DC-SQUID, one which gives rise to the washboard potential 
$E_{b}\gamma $ term as well as intrinsically irreversible dynamics. 
This irreversiblity arises as a direct consequence of the measurement process, and 
the starting point for our model is the master equation
\begin{equation}
\overset{.}{\rho }=-\frac{i}{\hbar }[ H,\rho ] +L \rho  L^{\dagger
}-\frac{L^{\dagger }L \rho }{2}-\frac{\rho  L^{\dagger } L}{2}, \label{TheLindbladEquation}
\end{equation}
where for $I_{b i a s}>0$ we have defined $L=\sqrt{|E_{b}|/\hbar }\eta ^{\dagger
}$ (see reference\cite{spiller}) and the charge-tunnelling non-unitary
operator on the large Josephson junction is $\eta ^{\dagger }\left|n\right\rangle  =\left|n+1\right\rangle $
and $\eta \left|n\right\rangle  =\left|n-1\right\rangle $. The state $|n\rangle $ represents the number of
Cooper pairs that have tunnelled through the large Josephson junction
(i.e an eigenstate of the Cooper pair number operator $N$) and the cooper pair tunnelling operator $\eta $ satisfies
$\left[ N,\eta ^{\dagger }\right] =\eta ^{\dagger }$ and $ \left[ N,\eta \right] =-\eta $. For $I_{bias}<0$ we have defined $L=\sqrt{|E_{b}|/\hbar
}\eta $. These Lindblad operators account for the movement of Cooper pairs across the Josephson
junction at an average rate given by the current $I_{b i a s}/2 e$.
That is, the operators $\eta ^{\dagger }$ and $\eta $ count the number of electrons added by the external bias current
to the large Josephson junction at an average rate $|E_{b}|/\hbar$. 

By introducing the Lindblad equation, given by Eq. (\ref{TheLindbladEquation}), we are proposing a heuristic method to model the bias current which attempts to capture the notion that the current source counts the number of electrons tunnelling through the Josephson junction. In the remainder of this section we reconcile such a model by showing that it is in fact in agreement  with a classical current biased Josephson junction and that the Lindblad terms contained in Eq. (\ref{TheLindbladEquation}) tend to destroy superpositions of different phase states.

Using Eq.~(\ref{TheLindbladEquation}), the master equation therefore reads
\begin{equation}
\overset{.}{\rho }=\left\{
\begin{array}{rl}
  -\frac{i}{\hbar }[ H,\rho ] +\frac{|E_{b}|}{\hbar
}\left( \eta ^{\dagger }\rho \eta -\rho \right) &    \text{  if  }
I_{bias}\geq 0 \\
  -\frac{i}{\hbar }[ H,\rho ] +\frac{|E_{b}|}{\hbar
}\left( \eta \rho \eta ^{\dagger }-\rho \right) &    \text{  if  }
I_{bias}<0.
\end{array}
\right. \label{XRef-Equation-830153234}
\end{equation}
In this equation the Hamiltonian $H$ describes
the Josephson junction or qubit dynamics but does not 
include the bias current washboard potential terms. 
Describing the current bias in superconducting circuits 
through this Lindblad superoperator is compatible 
with the phenomenology of the current biased Josephson 
junction in the classical limit $(C\rightarrow \infty )$ where 
the phase across the Josephson junction is fixed by the 
applied current. For instance, we can consider a single 
Josephson junction which is current biased ($I_{b i a s}\geq 0$) and described by 
Eq. (\ref{XRef-Equation-830153234}) where the Hamiltonian 
is given by
\begin{equation}
H=\frac{2e^{2}}{C}N^{2}+\frac{\Phi _{0}I_{C}}{2\pi }
\left( 1-\cos{\gamma}\right) \nonumber
\end{equation}
and $N$ is the Cooper pair number operator on the Josephson
junction.  In the steady state of this equation for the single 
Josephson junction, $\overset{.}{\rho }=0$,  we can compute the quantity
\begin{equation}
\frac{d\left\langle  N\right\rangle  }{d t}=\operatorname{Tr}( \overset{.}{\rho
}N) =0\nonumber,
\end{equation}
to look at the role of the bias current in our model. Using the cyclic property of the trace we find that
\begin{equation}
-i \operatorname{Tr}(  \rho \left [N,H\right] ) +|E_{b}| \operatorname{Tr}(
\eta ^{\dagger }\rho  \left( N+1\right) \eta -\rho  N) =0\nonumber
\end{equation}
and from the commutation relations for $\eta$ and $N$,
together with the definition of the current operator
\begin{align}
I&=\frac{2\pi }{\Phi _{0}}\frac{\partial H}{\partial \gamma }=I_{C}\sin
\gamma \nonumber\\
&=I_{C} \left( \frac{\eta ^{\dagger }-\eta }{2i}\right) =
\frac{2i \pi }{\Phi _{0}} \left[N,H\right] ,
\label{XRef-Equation-98122117}
\end{align}
we therefore show for $I_{b i a s}\geq 0$ that we have the expected result in the classical limit $(C\rightarrow \infty )$; that is
\begin{equation}
\left\langle  I\right\rangle  =\operatorname{Tr}( \rho  I) =I_{b i a s}.\nonumber
\end{equation}
Also by considering an oppositely biased current ($I_{b i a s}< 0$) 
we find that by the inclusion of the Lindblad terms for the bias current in the master
equation (Eq. (\ref{XRef-Equation-830153234})) we have 
$\left\langle  I\right\rangle =I_{b i a s}$ and therefore retained
the expected behaviour of the the bias current in the single Josephson
junction system. That is, the current through the Josephson junction
in our model is that applied by the current source. 

Additionally, in our model, the linear washboard term $E_{b}\gamma$
arises naturally from the Lindblad superoperator description of the bias current. 
By expanding the Lindblad superoperators in terms of its phase representation,
then to first order, we can obtain the reversible dynamics of the washboard potential 
through the $E_{b}\gamma$  term. The higher-order terms from this expansion provide
us with the intrinsic irreversible terms of the current source in our model. Hence 
having introduced the current source as an irreversible
one, the system can be approximated by the washboard potential
model of a current biased Josephson junction with an added irreversibility. 
For instance, by making an approximation to the full master equation 
(Eq. (\ref{XRef-Equation-830153234})) for the Quantronium circuit we 
can write the operators $\eta ^{\dagger}$ and $\eta  $ in their phase 
representation and approximate them
to second order. That is we can write
\begin{equation}
\label{XRef-Equation-830153453}
\eta ^{\dagger }=e^{+i\gamma }\approx 1+i \gamma
-\frac{\gamma ^{2}}{2}
\end{equation}
and
\begin{equation}
\label{XRef-Equation-83015354}
\eta =e^{-i\gamma }\approx 1-i \gamma -\frac{\gamma
^{2}}{2}.
\end{equation}
Under this approximation, and considering the cases for
$I_{b i a s}$ being positive and negative, the master equation of
the system is
\begin{equation}
\overset{.}{\rho }=-\frac{i}{\hbar }[ H-E_{b}\gamma ,\rho ]
+\frac{|E_{b}|}{\hbar }\left( \gamma \rho \gamma -\frac{\gamma ^{2}\rho
}{2}-\frac{\rho \gamma ^{2}}{2}\right) . \label{ApproxMasterEquation}
\end{equation}
Here we emphasise that the first order approximation to
the operator $L$ is the $E_{b}\gamma $ term which appears in the
tilted washboard potential model. The additional three terms appearing
at the end of the master equation are the irreversible decohering
terms of this model under our second order approximation.

By making the second order approximations (Eq. (\ref{XRef-Equation-830153453})
and Eq. (\ref{XRef-Equation-83015354})) for the operators $\eta
^{\dagger }$ and $\eta $, we have expanded them in terms of the operator
$\gamma $ about the point $\langle \gamma \rangle =0$; in this expansion
we have used the small parameter $\Delta $ which is the variance
of a sharply peaked Gaussian state in the phase representation.
For instance if we consider only the decoherence term in the $I_{b
i a s}>0$ master equation (Eq. (\ref{XRef-Equation-830153234})) we
have
\begin{equation}
\overset{.}{\rho }=\frac{|E_{b}|}{\hbar }\left( e^{+i \gamma }\rho
e^{-i \gamma }-\rho \right) =\frac{|E_{b}|}{\hbar }{\mathcal D}[
\gamma ] \rho .
\label{XRef-Equation-83015385}
\end{equation}
The steady state of this master equation can be written
as $\rho _{0}=|\psi _{0}\rangle \langle \psi _{0}|$ where $|\psi
_{0}\rangle $ is the sharply peaked Gaussian steady state wavefunction.
This wavefunction results from the small charging energy relative
to the Josephson energy of the junction. Note that this wavefunction, tightly peaked around a given value of $\gamma$, is consistent with the Josephson relation for a classical current passing through a Josephson junction. Therefore we write the
steady state wave function as 
\begin{equation}
 \left|\psi _{0}\right\rangle  =\frac{1}{\sqrt[4]{2 \pi
\Delta }}\int e^{-\gamma ^{2}/4\Delta }\left|\gamma \right\rangle  d
\gamma ,\nonumber
\end{equation}
where the variance $\Delta $ is small so that the wavefunction
is sharply peaked in phase. Using this wavefunction to construct
the steady state density matrix $\rho _{0}$ we can approximate the
term ${\mathcal D}[ \gamma ] \rho _{0}$ in the master equation (Eq.
(\ref{XRef-Equation-83015385})) as follows: 
\begin{align}
 {\mathcal D}[ \gamma ] \rho _{0} =&\frac{1}{\sqrt{2\pi
\Delta }}\int e^{i( \gamma -\gamma ^{\prime }) -\frac{\gamma ^{2}+\gamma
^{\prime 2}}{4\Delta }} \left|\gamma \right\rangle  \left\langle  \gamma
'\right|d\gamma 'd\gamma  
-\left|\psi _{0}\right\rangle  \left\langle  \psi
_{0}|\right.  \nonumber\\
  =&- i \sqrt{\Delta }\left[-\overset{\_}{\gamma },\rho _{0}\right]
-\Delta \left[\overset{\_}{\gamma },\left[ \overset{\_}{\gamma },\rho \right]
\right]+O\left [\Delta^{3/2}\right] ,\nonumber
\end{align}
where we have used the scaled phase operator 
$\gamma=\sqrt{\Delta }\overset{\_}{\gamma}$, and we have approximated
the exponential by its Taylor series expanded in terms of the small
parameter $\sqrt{\Delta }$. After also considering the case $I_{bias}<0$,
 we approximate the Lindblad derived decoherence
term in the master equation as 
\begin{equation}
 \overset{.}{\rho } =- i \sqrt{\Delta }\frac{E_{b}}{ \hbar}
 \left[-\overset{\_}{\gamma },\rho \right] -\Delta  \frac{|E_{b}|}{\hbar
} \left[\overset{\_}{\gamma },\left[ \overset{\_}{\gamma },\rho \right]
\right] +O\left [\Delta^{3/2}\right] \nonumber
\end{equation}
so that in the limit that $E_{b}\rightarrow \infty $ and
$\sqrt{\Delta }\rightarrow 0$ then $\sqrt{\Delta } E_{b}/\hbar $
is a constant $\widetilde{E}_{b}/\hbar $. In this limit the master
equation is 
\begin{equation}
\overset{.}{\rho }=-\frac{i}{\hbar }[ -\widetilde{E_{b}}\overset{\_}{\gamma
},\rho ] ,\nonumber
\end{equation}
which is the washboard potential arising from the bias-current. 
Thus the tilted washboard term arises naturally from our master equation,
accompanied by an intrinsically irreversible part.

\section{The Quantronium Measurement Model}

\begin{figure}[hbtp]
\begin{center}
\includegraphics[scale=0.5]{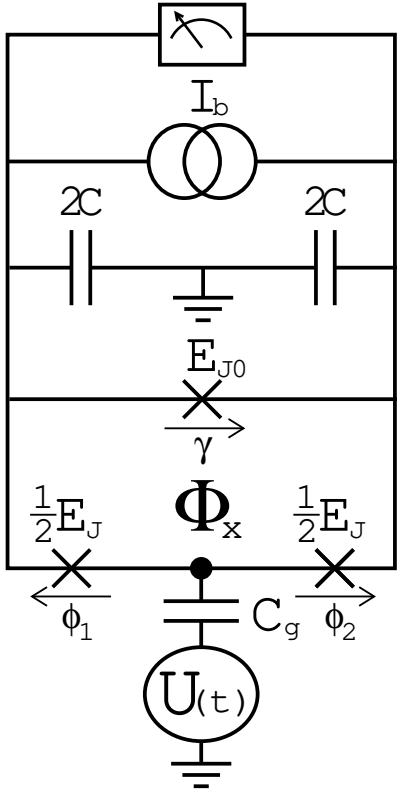}
\end{center}
\caption{Circuit diagram of the Quantronium qubit.}
\label{XRef-FigureCaption-830154719}
\end{figure}

In order to apply our irreversible current source approach to recent
experiments we consider the Quantronium qubit system depicted in
Fig.~\ref{XRef-FigureCaption-830154719}. The design of this charge-phase
qubit is similar to that of the Cooper pair box transistor\cite{JoyezPRL72-2458}.
The device consists of two identical low capacitance Josephson junctions
with a coupling energy $E_{J}/2$ and capacitance $C_{J}/2$. These
junctions are on either side of the isolated superconducting charge
`island' which is in a state of paired electron charge $2e N$, where
$N$ is the number of Cooper pairs on the island. This island is
incorporated into a superconducting loop with a larger Josephson
junction, which by design, has a coupling energy of $E_{J0}\sim
20E_{J}$ and a large shunt capacitance $C$; which was used in the
experiment to reduce phase fluctuations. The design of this device
requires that the characteristic energies $E_{J}$ and the charging
energy $E_{C}=2e^{2}/(C_{J}+C_{g}^{\prime })$, where $1/C_{g}^{\prime
}=1/C_{g}+1/4C$, are comparable so that neither charge or quantised
flux states in the loop are good quantum numbers. The discrete energy
states of the device are quantum superpositions of several charge
states\cite{{cottet, cottetThesis}}. Control of the qubit is made
via the pulsed microwave voltage source $U( t) $ which is capacitively
coupled to the Cooper pair box by the capacitor $C_{g}$, and the
applied flux $\Phi _{x}$ through the three junction superconducting
loop. These provide the elementary single qubit manipulations.

For this device the relation $\delta =\gamma +2e\Phi _{x}/\hbar
$ between the combined phase $\delta =\phi _{1}-\phi _{2}$ across
the Josephson junctions of the Cooper pair box, and the phase $\gamma
$ across the larger Josephson junction provides the readout process
of the Quantronium quantum state. From this relation the two lowest
energy states of the Quantronium have different persistent currents
in the three junction loop. This difference is used for state readout,
a current pulse $I_{bias}( t) $ from the `ideal' current source
is applied where the height of the pulse is chosen so that the transition
to a voltage state is made for only one of the Quantronium energy
eigenstates; when the addition of the loop persistent current state
and the current pulse exceeds the critical current of the large
junction. This process discriminates between the two qubit states
associated with the two lowest levels of the Quantronium.

The Hamiltonian for the Quantronium, which we consider in terms of 
the master equation Eq. (\ref{XRef-Equation-830153234}) and its 
approximation Eq. (\ref{ApproxMasterEquation}), that is without 
the energy term corresponding to the readout current source 
$I_{bias}$ is 
\begin{align}
H=&\phantom{+}E_{C}( N-N_{g}) ^{2}+E_{J}\left( 1-\cos \left( \frac{\phi +\gamma }{2}\right)
\cos{\varphi }\right)\nonumber \\ &+\frac{Q^{2}}{2 C}+E_{J0}\left( 1-\cos \gamma
\right) .\nonumber
\end{align}
Here we have used the terms: the phase operator $\varphi =(\phi _{1}+\phi _{2})/2$
which is conjugate to the Cooper pair number operator
$N$, the dimensionless gate charge $N_{g}=C_{g}U/2e$, the phase bias
$\phi =2 \pi  \Phi _{x}/\Phi _{0}$ where $\Phi _{0}=h/2e$ is the
flux quantum, and the charge $Q$ on the large Josephson junction
with shunt capacitance $C$. In this Hamiltonian we have neglected
the energy term corresponding to the loop inductance of the device
based on the size of the device.

To analyse the measurement induced decoherence in our model we simplify
the Hamiltonian $H$ by considering the dynamics of the lowest two qubit eigenstates
where we use $|0\rangle $ and $|1\rangle $ to denote
the lowest and first excited state of the Quantronium system respectively.
Here we work at the point where the applied flux $\Phi _{x}$ is
set to 0 and the dimensionless gate charge $N_{g}$ is set to $1/2$.
In this configuration the qubit energy levels are separated by the Josephson
Junction coupling energy so we write our Hamiltonian as
\begin{align}
H^{\prime }=&\phantom{+}E_{J}\left( 1-\cos \left( \frac{\gamma }{2}\right) \right) \sigma _{z}+E_{C0}N^{2}\nonumber\\
&+E_{J0}\left(1-\cos \gamma \right) ,
\label{XRef-Equation-830155253}
\end{align}
where $E_{C0}=2e^{2}/C$ and $N$ is the charge operator
for the large Josephson junction which is conjugate to the phase
operator $\gamma $. This Hamiltonian describes a two level system
separated by an energy $E_{J}( 1-\cos ( \gamma /2) ) $ where the
phase $\gamma $ provides the coupling between the qubit and the
readout junction. If we now assume that the large junction is in a localised semi-classical
state near $\langle \gamma \rangle =0$, then by expanding $\cos\gamma$ 
in Eq. (\ref{XRef-Equation-830155253}) to second order in $\gamma$, 
we obtain the Hamiltonian
\begin{equation}
H_{R}=E_{C0}N^{2}+\frac{E_{J0}\gamma ^{2}}{2}+\frac{E_{J}\gamma
^{2}\sigma _{z}}{8},
\label{XRef-Equation-830155342}
\end{equation}
which has a form of a displaced simple harmonic oscillator.

\section{The DC-SQUID Measurement Model}

\begin{figure}[hbtp]
\begin{center}
\includegraphics[scale=0.5]{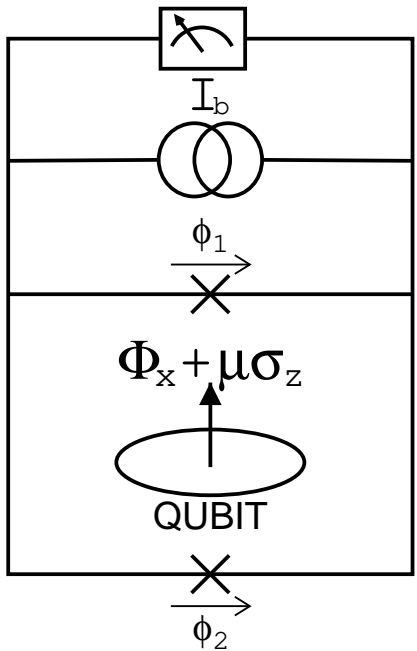}
\end{center}
\caption{Circuit diagram of a DC-SQUID used for qubit state detection
via a measurement of the qubit's magnetic flux.}
\label{XRef-FigureCaption-830155225}
\end{figure}

In addition to the Quantronium experiments our approach 
is applicable to the systems where a two level quantum
device has been measured by a small inductance DC-SQUID such as
the persistent current qubit\cite{delft}. In these experiments the
coherent oscillations in a low inductance three Josephson junction
qubit structure have been observed. Similarly, the use of low inductance
microSQUID\cite{MicroSQUID} structures have been proposed to readout
the quantum state of nanometre scale magnetic particles of large
spin and high anisotropy molecular clusters\cite{spiller2}. Here 
the measurement of a magnetic flux quantum state inductively
coupled to a DC-SQUID with a low inductance relies
on the induced change of the effective critical current of the the
DC-SQUID, for this type of measurement a current ramp scheme is
used which is similar to that used in the Quantronium readout process.

In Fig.~\ref{XRef-FigureCaption-830155225} we consider two Josephson
junctions with a coupling strength $E_{J0}/2$, capacitance $C_{J0}/2$
and phases (as shown) of $\phi _{1}$ and $\phi _{2}$ in a superconducting
loop. For this device we define the total phase $\gamma =(\phi _{1}+\phi
_{2})/2$ across the device and the applied flux
\begin{equation}
2\pi \left( \frac{\Phi _{x}-\mu \,\sigma _{z}}{\Phi _{0}}\right) =\phi _{1}-\phi
_{2} \nonumber
\end{equation}
where $\mu \sigma _{z}$ is the magnetic flux of the qubit
state. When the loop inductance is small then
the flux through the loop $\Phi \approx \Phi _{x}-\mu\, \sigma
_{z}$, also when the charging energy of the Josephson junctions is small so
that the quantum state of the DC-SQUID detector is well defined
in phase and the energy of the first excited state of the detector
is larger than the other energies of the system so that it exhibits
ground state behaviour, then we can write the Hamiltonian of the system
for the master equation Eq. (\ref{XRef-Equation-830153234}) and its 
approximation Eq. (\ref{ApproxMasterEquation}) as 
\begin{equation}
H=H_{Q}+H_{S},\nonumber
\end{equation}
here the DC-SQUID Hamiltonian is 
\begin{equation}
H_{S}=\frac{\left( e\,N\right) ^{2}}{2C_{J0}}+E_{J0}\, ( 1-\cos \gamma
\,\cos ( \varphi _{x}-\delta \varphi\, \sigma _{z}) ) \nonumber
\end{equation}
where $\varphi _{x}-\delta \varphi \,\sigma _{z}=\pi ( \Phi
_{x}-\mu\, \sigma _{z}) /\Phi _{0}$ and the qubit Hamiltonian is 
$H_{Q}=(\epsilon _{0}\,\sigma _{z}+t_{0}\,\sigma _{x})/2.$ For 
small $\delta \varphi $, eliminating the constant
terms and assuming that the tunnelling between the flux
states of the qubit has been turned off, $t_{0}=0$, we simplify this Hamiltonian $H$
to
\begin{align}
H^{\prime }=&\phantom{-}\frac{\epsilon _{0}}{2}\,\sigma _{z}+E_{C0}\, N^{2}-E_{J0}\,\cos
\varphi _{x}\,\cos{\gamma }\nonumber \\&-E_{J0}\,\delta \varphi\, \sigma
_{z}\,\sin \varphi _{x}\,\cos{\gamma }. \nonumber
\end{align}
This Hamiltonian is similar to Eq. (\ref{XRef-Equation-830155253}),
and since we assume that the DC-SQUID is localised near
$\langle \gamma \rangle =0$ we again make a second order approximation
to the $\gamma $ terms to arrive at the reduced Hamiltonian 
\begin{equation}
H_{\mathrm{R}}=E_{C0}\, N^{2}+\frac{E_{J0}\,\cos \varphi _{x}\,\gamma ^{2}}{2}+\frac{E_{J0}\,\delta
\varphi \,\sin \varphi _{x}\,\gamma ^{2}\,\sigma _{z}}{2},\nonumber
\end{equation}
when the qubit energy level separation satisfies $\epsilon
_{0}/2=E_{J0}\,\delta \varphi \,\sin \varphi _{x}$. Since the form of
this Hamiltonian is identical to Eq. (\ref{XRef-Equation-830155342})
then the model presented for the Quantronium can be directly applied
to the measurement of the magnetic flux of a qubit with a low inductance
DC-SQUID.

\section{The Reversible Current Source}
\label{XRef-Section-102193751}
\subsection{The Reversible Current Source Wigner Function }
\label{XRef-Subsection-102193736}
To investigate the Hamiltonian dynamics of the
Quantronium measurement model (and by analogy the DC-SQUID measurement
model) we consider the simplified Quantronium Hamiltonian derived
in the previous section: 
\begin{equation}
H_{R}=E_{C0}\, N^{2}+\frac{E_{J0}\, \gamma ^{2}}{2}+\frac{E_{J}\, \gamma
^{2}\, \sigma _{z}}{8}.
\label{XRef-Equation-830155424}
\end{equation}
We use this Hamiltonian to analyse the measurement induced decoherence
relative to the washboard potential phenomenology, which does not
include the effects of decoherence. In this section we derive 
the decoherence-free dynamics of the system
using the standard tilted-washboard model by including the term 
$E_{b}\gamma $ in the Hamiltonian Eq. (\ref{XRef-Equation-830155424}).
In this model, the density matrix for the qubit and the readout device evolves
according to
\begin{equation}
\overset{.}{\rho }=-\frac{i}{\hbar }[ H_{R}-E_{b}\gamma ,\rho ].
\end{equation}
We decompose $\rho$ as
\begin{align}
 \rho =&\phantom{+}\rho _{+}( t) \otimes \left|0\right\rangle  \left\langle
0\right|+\rho _{\times }( t) \otimes \left|0\right\rangle  \left\langle  1\right|\nonumber \\ &+{\rho
_{\times }}^{\dagger}( t) \otimes \left|1\right\rangle  \left\langle  0\right|+\rho
_{-}( t) \otimes \left|1\right\rangle  \left\langle  1\right|, 
\label{XRef-Equation-911164614}
\end{align}
where $\rho _{+}$ and $\rho _{-}$ describe the evolution
of the Josephson junction when the qubit is in the states $|0\rangle $
and $|1\rangle $ whilst $\rho _{\times }$ describes the coherence
between them. We assume the initial state of the system $\rho ( 0)
$ is a product state of the readout Josephson junction density matrix
$w$ and the qubit in the symmetric state $(|0\rangle +|1\rangle
)/\sqrt{2}$, so
\begin{equation}
\rho ( 0) =\frac{w( 0) }{2}\otimes \left( |0\right\rangle  \left\langle
0|+|0\right\rangle  \left\langle  1|+|1\right\rangle  \left\langle
0|+|1\right\rangle  \left\langle  1|\right) .\nonumber
\end{equation}

The dynamics of $\rho _{+}$ and $\rho_{-}$ do not depend on $\rho_{\times}$,
so their dynamics are described by the qubit-state dependent Hamiltonian
\begin{equation}
H_{R\pm }=E_{C0}N^{2}+\frac{E_{J0}\gamma ^{2}}{2}\pm \frac{E_{J}\gamma
^{2}}{8}.\nonumber
\end{equation}
We define the two sets of raising and lowering operators $a_{\pm
}^{\dagger }$ and $a_{\pm }$ by
\begin{equation}
\gamma =\Gamma _{\pm }( a_{\pm }^{\dagger }+a_{\pm }) =\sqrt{\frac{\lambda
_{\pm }}{2}}\left( a_{\pm }^{\dagger }+a_{\pm }\right) 
\label{XRef-Equation-910164742}
\end{equation}
and 
\begin{equation}
N=-\frac{i}{\sqrt{2\lambda _{\pm }}}\left( a_{\pm }-a_{\pm }^{\dagger
}\right) ,
\label{XRef-Equation-910164815}
\end{equation}
where $\lambda _{\pm }=\sqrt{2\nu /{(1\pm \mu /4)}}$, $\nu =E_{C0}/E_{J0}$, $\mu =E_{J}/E_{J0}$,
and $H_{R\pm }=\hbar \omega _{\pm}a_{\pm }^{\dagger }a_{\pm }$ where 
\begin{equation}
\hbar \omega _{\pm }=\sqrt{2E_{C0}E_{J0} \left( 1\pm \frac{E_{J}}{4E_{J0}}\right)
}.\nonumber
\end{equation}
Using these scalings we can define the two independent
equations for $\rho _{+}$ and $\rho _{-}$ 
\begin{align}
\label{XRef-Equation-92216732}
\overset{.}{\rho }_{+}&=-i\left[ \omega _{+}a_{+}^{\dagger }a_{+}-\frac{E_{b}\Gamma
_{+}}{\hbar }\left( a_{+}^{\dagger }+a_{+}\right) ,\rho _{+}\right]  \\
\overset{.}{\rho }_{-}&=-i\left[ \omega _{-}a_{-}^{\dagger }a_{-}-\frac{E_{b}\Gamma
_{-}}{\hbar }\left( a_{-}^{\dagger }+a_{-}\right) ,\rho _{-}\right] .
\end{align}
By the anti-commutation relation, $\{ A,B\} = AB+BA$, we define the 
equation for the off-diagonal element $\rho _{\times }$ as
\begin{align}
 \overset{.}{\rho }_{\times }  =&-i\left[ \omega _{\times }a_{\times
}^{\dagger }a_{\times }-\frac{E_{b}\Gamma _{\times }}{\hbar }\left(
a_{\times }^{\dagger }+a_{\times }\right) ,\rho _{\times }\right] \nonumber \\
&-i\frac{E_{J}{\Gamma_{\times }}^{2}}{8\hbar }\left\{ \left( a_{\times }^{\dagger }+a_{\times
}\right) ^{2},\rho _{\times }\right\}.
\label{XRef-Equation-830155523}
\end{align}
For the off-diagonal component $\rho _{\times }$ we have defined the raising and lowering
operators $a_{\times }^{\dagger }$ and $a_{\times }$ where 
\begin{gather}
\gamma =\Gamma _{\times }( a_{\times }^{\dagger }+a_{\times }) =\sqrt{\frac{\lambda
_{\times }}{2}}\left( a_{\times }^{\dagger }+a_{\times }\right),
\label{XRef-Equation-102193326}
\\N=-\frac{i}{\sqrt{2\lambda _{\times }}}\left( a_{\times }-a_{\times
}^{\dagger }\right) ,
\end{gather}
$\lambda _{\times }=\sqrt{2\nu }$, and $\hbar \omega _{\times }=\sqrt{2E_{C0}E_{J0}}.$
The master equation for $\rho_{\times}$ defines the dynamics of 
both the off-diagonal elements of the density
matrix, where the equation for ${\rho _{\times }}^{*}$ is the Hermitian
conjugate of Eq. (\ref{XRef-Equation-830155523}).

To solve the dynamics of the system, we transform to a Wigner 
representation of the state\cite{{wigner, milburn}}. To obtain the 
equation of motion for the Wigner function we first derive the characteristic
function equation of motion $\partial \Upsilon (\beta )/\partial
t=\operatorname{Tr}( D\overset{.}{\rho }) $, where the characteristic
function is defined as $\Upsilon ( \beta ) =\operatorname{Tr}( \operatorname{D\rho
}) $ and $D$ is the displacement operator defined by $D=\exp ( \beta
a^{\dagger }-\beta ^{*}a) $. Writing $D$ in normal and anti-normal
order then we can find the relevant operator rules for converting
to the characteristic function equations. The Wigner function equation of motion is found by
taking the Fourier transform of the characteristic 
function equation of motion $\overset{.}{\Upsilon }( \beta ) $. Thus
\begin{equation}
\overset{.}{W}( \alpha ) =\int _{-\infty }^{+\infty }e^{\beta ^{*}
\alpha -\beta  \alpha ^{*}}\overset{.}{\Upsilon }( \beta ) d^{2}\beta .
\label{XRef-Equation-102193535}
\end{equation}
After performing this procedure we use a compact notation to write down the Wigner
function equations from the three master equations Eq. (\ref{XRef-Equation-92216732})
- Eq. (\ref{XRef-Equation-830155523}); for the operators $a_{+}$,
$a_{-}$ and $a_{\times }$ defined in the three master equations we
correspondingly have the complex parameters $\alpha _{+}$, $\alpha
_{-}$ and $\alpha _{\times }$ but we drop the subscripts since they
appear separately in the three characteristic function equations.
This procedure provides us with three uncoupled equations 
\begin{align}
\overset{.}{W}_{+}( \alpha ) =&-i\omega _{+}( \partial _{\alpha ^{*}}\alpha
^{*}-\partial _{\alpha }\alpha ) W_{+}( \alpha )\nonumber \\ &-\frac{i E_{b}\Gamma
_{+}}{\hbar }\left( \partial _{\alpha }-\partial _{\alpha ^{*}}\right)
W_{+}( \alpha ) ,
\label{XRef-Equation-910121447} \\
\overset{.}{W}_{-}( \alpha ) =&-i\omega _{-}( \partial _{\alpha
^{*}}\alpha ^{*}-\partial _{\alpha }\alpha ) W_{-}( \alpha )\nonumber \\ &-\frac{i
E_{b}\Gamma _{-}}{\hbar }\left( \partial _{\alpha }-\partial _{\alpha
^{*}}\right) W_{-}( \alpha ) ,
\label{XRef-Equation-91012152} \\
\nonumber\overset{.}{W}_{\times }( \alpha ) =&
-i\omega _{\times }( \partial _{\alpha ^{*}}\alpha ^{*}-\partial
_{\alpha }\alpha ) W_{\times }( \alpha )\\ \nonumber &-\frac{i E_{b}\Gamma _{\times }}{\hbar }\left( \partial
_{\alpha }-\partial _{\alpha ^{*}}\right) W_{\times }( \alpha )\\ 
&-\frac{i E_{J}{\Gamma _{\times}}^{2}}{8 \hbar }\left( 2\left( \alpha ^{*}\right) ^{2}+2\alpha ^{2}+4|\alpha
|^{2}\right) W_{\times }( \alpha )\nonumber\\ &
-\frac{i E_{J}{\Gamma _{\times}}^{2}}{16 \hbar }\left( \partial _{\alpha ^{*}}^{2}+\partial _{\alpha
}^{2}-2\partial _{\alpha }\partial _{\alpha ^{*}}\right) W_{\times
}( \alpha ).\label{XRef-Equation-92917409}
\end{align}
We note that each of these three equations are described in
three separate co-ordinate spaces related to each other by a small
scaling factor. This same procedure will be used in the description
of the irreversible current source described in the following section.
Eq. (\ref{XRef-Equation-92917409}) can be expressed in terms of the 
phase $\gamma$ and charge $N$ variables using the definitions 
Eq. (\ref{XRef-Equation-910164742}) and 
Eq. (\ref{XRef-Equation-910164815}), doing so we find
\begin{align}
 \nonumber\overset{.}{W}_{\times }( \gamma ,N) =&-\omega _{\times }\left( \lambda
_{\times }\partial _{\gamma }N-\frac{1}{\lambda _{\times }}\partial
_{N}\gamma \right) W_{\times }( \gamma ,N)\\ \nonumber  &-\frac{E_{b}}{\hbar }\partial
_{N}W_{\times }( \gamma ,N) \nonumber\\ &-\frac{iE_{J}}{16 \hbar }\left( 4\gamma
^{2}-\partial _{N}^{2}\right) W_{\times }( \gamma ,N) .
\label{XRef-Equation-10114363}
\end{align}

\subsection{The Reversible Current Source Wigner Function Solution}

The first two Wigner function equations (Eq. (\ref{XRef-Equation-910121447})
and Eq. (\ref{XRef-Equation-91012152})) for the readout Josephson
junction density matrix component elements $\rho _{+}$ and $\rho
_{-}$ can be solved analytically using the Wang and Uhlenbeck solution
for a linear Fokker-Plank equation\cite{WangMilburn} since the equations
are of the form
\begin{equation}
\overset{.}{W}_{\pm }( \alpha _{\pm },t) =\left( -\nabla _{z}^{T}.M_{\pm
}.z+\nabla _{z}^{T}.N_{\pm }.\nabla _{z}/2\right) W_{\pm }( \alpha
_{\pm },t) .
\label{XRef-Equation-102194049}
\end{equation}
where
\begin{gather}
M_{\pm }=\left( \begin{array}{cc}
 -i \omega _{\pm } & 0 \\
 0 & i \omega _{\pm }
\end{array}\right) ,N_{\pm }=0,\nonumber \\\nabla _{z}=\binom{\partial _{\widetilde{\alpha
}}}{\partial _{\widetilde{\alpha }^{*}}},z=\binom{\widetilde{\alpha
}}{\widetilde{\alpha }^{*}},\nonumber
\end{gather}
and $\widetilde{\alpha }=\alpha -E_{b} \Gamma _{\pm }/(\hbar
\omega _{\pm })$. From these two solutions $W_{+}$ and $W_{-}$ we
can specify the Wigner function for the reduced state of the readout
junction, since from the definition of the Wigner function we have
\begin{align}
W( \gamma ,N,t) =&\int _{-\infty }^{+\infty }e^{\beta ^{*} \alpha
-\beta  \alpha ^{*}} \operatorname{Tr}( D \rho ) d^{2}\beta \nonumber \\
=&\frac{1}{2}\left(
W_{+} +W_{-} \right) ,
\label{XRef-Equation-92217507}
\end{align}
where the trace is performed over the Josephson junction
and qubit states. From the Wigner function we can obtain a probability
distribution for the state of the system in the state variables
$\gamma $ or $N$ by integrating over the state variable for the
state variable $N$ or $\gamma $ respectively. 

This Wigner function for the combined system 
does not show the coherence that exists between the 
states of the qubit, that is it cannot be used to distinguish 
between a pure and a mixed state. We therefore
construct a function from the three equations for the Wigner function
terms $W_{+}\text{}$, $W_{-}$ and $W_{\times }$ that we derived
from the readout Josephson junction density matrix component elements
$\rho _{+}$, $\rho _{-}$, and $\rho _{\times }$; this function is
found by directly Wigner transforming both sides of Eq. (\ref{XRef-Equation-911164614})
over the Josephson junction degrees of freedom which defines the
operator
\begin{align}
\hat{W}_{s}( \gamma ,N,t) =&\phantom{+}W_{+}|0\left\rangle  \right\langle
0|+W_{-} |1\left\rangle  \right\langle 1|\nonumber\\
&+{W_{\times }}^{*} |1\left\rangle  \right\langle
0|+W_{\times } |0\left\rangle  \right\langle
1|.\nonumber
\end{align}
From this we can calculate the projection onto the initial
state $W_{s}( \gamma ,N,t) =\left\langle +\right|\hat{W}_{s}( \gamma ,N,t)
\left|+\right\rangle $ where $\left|+\right\rangle =\left(\left|1\right\rangle +\left|0\right\rangle \right)/\sqrt{2}$
so that
\begin{equation}
W_{s}( \gamma ,N,t) =\frac{1}{2}\left( W_{+} +W_{-} \right) +\operatorname{Re}( W_{\times }
) .\nonumber
\end{equation}
Integrating this function over the canonical coordinates
gives the probability to find the system in the initial state at
time $t$.

The solutions for the diagonal Wigner function terms $W_{+}$ and
$W_{-}$ obtained from the Eq. (\ref{XRef-Equation-910121447}) and
Eq. (\ref{XRef-Equation-91012152}) are the Gaussians
\begin{equation}
W_{\pm }( \alpha _{\pm },t) =\frac{1}{2\pi |C_{\pm }|}\exp \left( -\frac{1}{2}{u_{\pm
} }^{T}.C_{\pm }^{-1}.u_{\pm }
\right) ,
\label{XRef-Equation-830155822}
\end{equation}
where 
\begin{equation}
u_{\pm }( \alpha _{\pm },t) =\left(
\begin{array}{c}
	 \alpha _{\pm }-\frac{E_{b}\Gamma_{\pm }}{\hbar  \omega _{\pm }}-
	 e^{-i \omega _{\pm }t}\left( \alpha _{0}-\frac{E_{b}\Gamma_{\pm }}{\hbar  \omega_{\pm }}\right) \\
	 \alpha _{\pm }^{*}-\frac{E_{b}\Gamma_{\pm }}{\hbar  \omega _{\pm }}-
	 e^{+i \omega _{\pm }t}\left( \alpha _{0}^{*}-\frac{E_{b}\Gamma_{\pm }}{\hbar  \omega _{\pm }}\right) \\
\end{array} \right)	\nonumber
\end{equation}
and the covariance matrix 
\begin{equation}
C_{\pm }=\left( \begin{array}{cc}
  e ^{-i \omega _{\pm } t} & 0 \\
 0 &  e ^{i \omega _{\pm }t}
\end{array}\right) .C_{0}.\left( \begin{array}{cc}
  e ^{-i \omega _{\pm } t} & 0 \\
 0 &  e ^{i \omega _{\pm }t}
\end{array}\right) \nonumber
\end{equation}
which decays from the initial covariance matrix: 
\begin{equation}
C_{0}=\left( \begin{array}{cc}
 \left\langle  \alpha ^{2}\rangle _{0}\right. -\left\langle  \alpha
\rangle _{0}\right. ^{2} & \left\langle  |\alpha |^{2}\rangle _{0}\right.
-|\left\langle  \alpha \rangle _{0}\right. |^{2} \\
 \left\langle  |\alpha |^{2}\rangle _{0}\right. -|\left\langle 
\alpha \rangle _{0}\right. |^{2} & \left\langle  \left( \alpha ^{*}\right)
^{2}\right\rangle  _{0}-\left\langle  \alpha ^{*}\rangle _{0}\right.
^{2}
\end{array}\right) \nonumber
\end{equation}
The solution (Eq. (\ref{XRef-Equation-830155822})) for
the terms $W_{+} $ and $W_{-}$ correspond to Gaussian functions in the $ (\alpha _{\pm },\alpha
_{\pm }^{*})$ co-ordinate space. The initial state of the Josephson
junction at $t=0$ is a Gaussian centred about zero, so that at the
instant the bias current $I_{bias}$ is applied 
\begin{align}
W_{s}( \alpha _{\times },0) =&\frac{2}{\pi }\exp \left( -2\alpha _{\times
}\alpha _{\times }^{*}\right)\nonumber \\ =&\frac{2}{\pi }\exp \left( -\frac{\gamma ^{2}}{\lambda
_{\times }}-\lambda _{\times } N^{2}\right) ,
\label{XRef-Equation-911221614}
\end{align}
this initial condition implies $\alpha_{0}=0$, and the covariance matrix
\begin{align}
C_{\pm }=C_{0}&=\frac{1}{4 \lambda _{\pm } \lambda _{\times }}\left(
\begin{array}{cc}
 {\lambda _{\times }}^{2}-{\lambda _{\pm }}^{2} & {\lambda _{\pm }}^{2}+{\lambda
_{\times }}^{2} \\
 {\lambda _{\pm }}^{2}+{\lambda _{\times }}^{2} & {\lambda _{\times }}^{2}-{\lambda
_{\pm }}^{2}
\end{array}\right) \nonumber \\ &\overset{\operatorname*{\textup{Lim}}\limits_{E_{\textup{J0}}\gg E_{J}}}{\longrightarrow
}\left( \begin{array}{cc}
 0 & 1/2 \\
 1/2 & 0
\end{array}\right) .
\label{XRef-Equation-102194935}
\end{align}
From the analytic solutions for the diagonal Wigner function terms
$W_{+}$ and $W_{-}$ we see that they correspond to fixed width Gaussian
curves that rotate on elliptical orbits through the $(\gamma ,N)$
co-ordinate space at different frequencies $\omega _{+}$ and $\omega
_{-}$ and the centre of the orbits are located at $E_{b}\Gamma _{+}/(\hbar
\omega _{+})$ and $E_{b}\Gamma _{-}/(\hbar  \omega _{-})$ along
the phase axis respectively. These diagonal Wigner function terms
in the absence of decoherence maintain their width and hence their
noise characteristics during their evolution. Hence the Wigner function
follows a complicated periodic motion. For instance, after a certain
number of oscillations at time $2T_{0}=2\pi /(\omega _{+}-\omega
_{-})$ the Wigner function term $W_{+}$ with the larger frequency
$\omega _{+}$ has completed an extra oscillation about its elliptical
orbit compared to the Wigner function term $W_{-}$. 

For the off-diagonal Wigner function term $W_{\times }( \gamma ,N,t) $ 
we assume a solution of the form
\begin{align}
W_{\times }( \gamma ,N,t) =\exp ( &a( t) \gamma +b( t) N+c( t) \gamma
^{2}\nonumber \\ &+d( t) \gamma N+e( t) N^{2}+f( t) ) .
\label{XRef-Equation-103122512}
\end{align}
From Eq. (\ref{XRef-Equation-10114363}), the coefficients in the exponent 
evolve according to
\begin{align}
\label{XRef-Equation-103123419}
\overset{.}{a}( t) &=\frac{\omega }{\lambda _{\times }}b( t) +G
d( t) +\frac{\left( -I+i E\right) }{2}b( t) d( t) ,
\\
\overset{.}{b}( t) &=-\lambda _{\times }\omega  a( t) +2G e( t)
+\left( -I+i E\right) e( t) b( t) ,
\\
\overset{.}{c}( t) &=\frac{\omega }{\lambda _{\times }}d( t) -i
E+\frac{\left( -I+i E\right) }{4}d( t) ^{2},
\\
\overset{.}{d}( t) &=-2\lambda _{\times }\omega  c( t) +\frac{2\omega
}{\lambda _{\times }}e( t) +\left( -I+i E\right) e( t) d( t) ,
\\
\overset{.}{e}( t) &=-\lambda _{\times }\omega  d( t) +\left( -I+i
E\right) e( t) ^{2},
\\ \label{XRef-Equation-103123443}
\overset{.}{f}( t) &=G b( t) +\frac{\left( -I+i E\right) }{4}\left(
2e( t) +b( t) ^{2}\right) ,
\end{align}
where $G=-E_{b}/\hbar $, $E=E_{J}/4 \hbar $ and $I=0$. Using
the initial conditions $a( 0) =0$, $b( 0) =0$, $c( 0) =-1/\lambda
_{\times }$, $d( 0) =0$, $e( 0) =-\lambda _{\times }$, $f( 0) =\ln
( 1/\pi ) $ we can solve for $W_{\times }( \gamma ,N,t) $ numerically. 

\begin{figure}[hbtp]
\begin{center}
\includegraphics{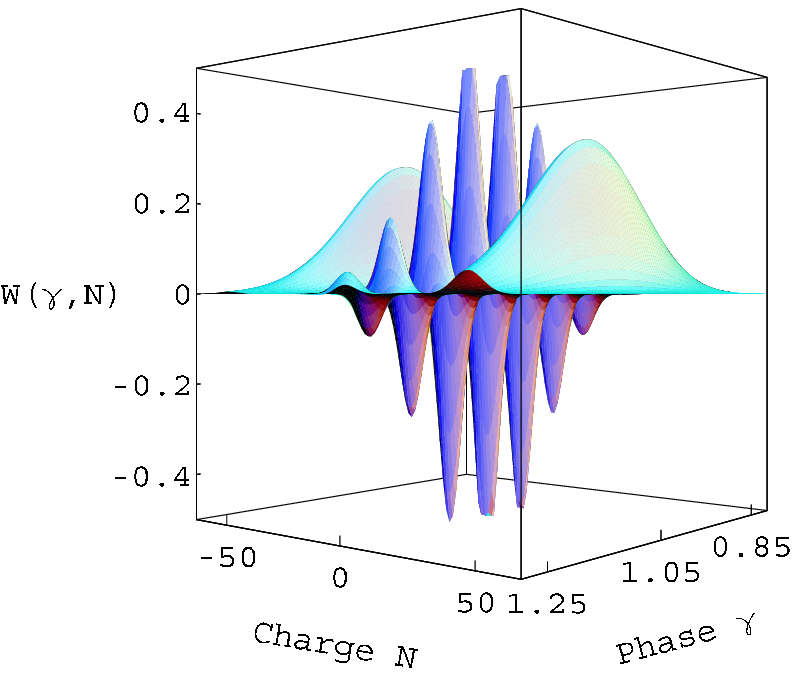}
\end{center}
\caption{Plot of the full Wigner Function $W_{s}( \gamma ,N,t) $
for the readout Josephson junction. The parameters used in this
plot are taken from the Quantronium experiment\cite{{Vion, cottetThesis}}
and they are: $E_{J}=0.86 k_{b} K$, $E_{C}=0.68 k_{b} K$, $\text{}E_{\textup{C0}}=0.0037
k_{b} K$, $E_{\textup{J0}}=18.4 k_{b} K$ and $E_{b}=0.97 E_{\textup{J0}}$
The time after which the bias current of $0.77 \mu A$ is applied
for this plot is $t= 16.932 \operatorname{ns}$. The plot demonstrates
the different terms that appear in the decoherence free Wigner function
for the readout Josephson junction. In the plot we can see that
at this particular time the two Gaussian curves corresponding to
the diagonal Wigner function terms $W_{+}$ and $W_{-}$ are separated
and the interference fringes that correspond to the off-diagonal
coherence term $\operatorname{Re}( W_{\times }) $ appears between
them.}
\label{XRef-FigureCaption-912183141}
\end{figure}

The plot of the full Wigner function $W_{s}( \gamma ,N,t) $ for the readout
Josephson junction is plotted using the experimental parameters
of the Quantronium experiment\cite{cottetThesis} in Fig.~\ref{XRef-FigureCaption-912183141}.
The interference fringes, centrally located between the 
two Gaussians $W_{+}$ and $W_{-}$, is due to 
the coherence between the qubit states, arising from $W_{\times}$.
The number of interference fringes present at a particular time $t$
is related to the separation of the two Gaussians in the $(\gamma ,N)$ 
co-ordinate space, which increases the further the terms $W_{+}$ and $W_{-}$ 
are apart. As the Gaussians separate the centre of the 
off-diagonal Wigner term $W_{\times }$ follows the trajectory 
shown in Fig.~\ref{XRef-FigureCaption-926175556}. In these figures we see that the main feature of these plots is
that the noise properties and the interference fringes are conserved
over time. In the absence of decoherence they continuously evolve
with a complicated periodic motion. This will be contrasted against
the evolution of the state in the presence of the irreversible
bias current decoherence in the next section.

\begin{figure}[hbtp]
\begin{center}
(a)\includegraphics[scale=0.83]{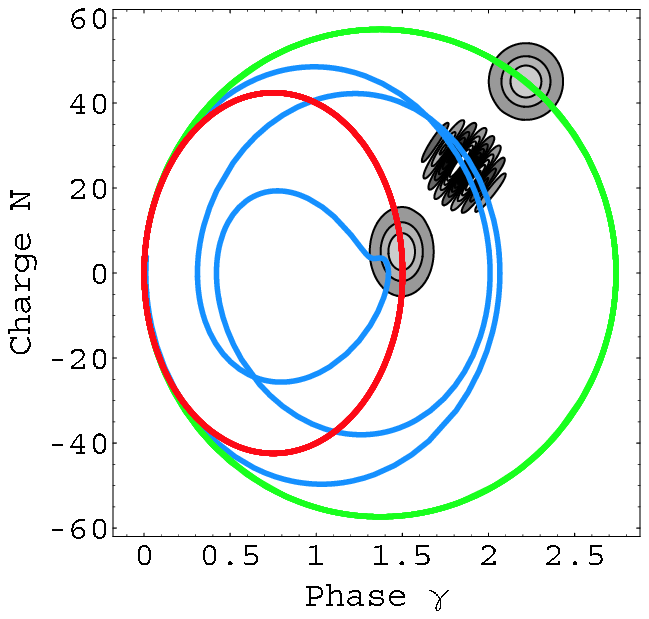}\\
(b)\includegraphics[scale=0.83]{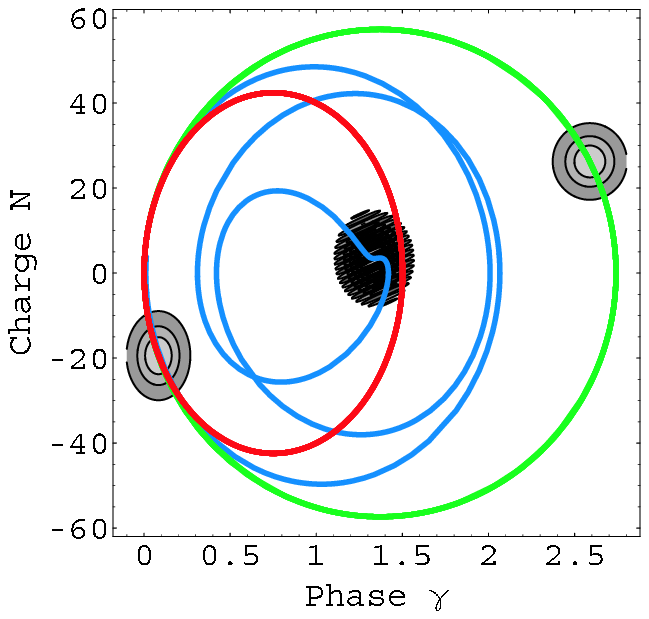}
\end{center}
\caption{Contour plots of the decoherence free Wigner function $W_{s}(
\gamma ,N,t) $ with the trajectories of the centre of the three
Wigner function components $W_{+}( \gamma ,N,t) $, $W_{-}( \gamma
,N,t) $ and $\operatorname{Re}( W_{\times }( \gamma ,N,t) ) $ superimposed
in red, green and blue respectively. (a) The contour plot of the
three Wigner function terms is for time $t=T_{0}/4$. (b) The contour
plot is for time $t=T_{0}$. Where the time $T_{0}$ is defined as
the time when the diagonal terms $W_{+}( \gamma ,N,t) $ and $W_{-}(
\gamma ,N,t) $ are the most separated and sit on opposite sides
of their respective trajectory ellipse, $T_{0}= \pi /(\omega _{+}-\omega
_{-})$. The parameters used in this plot are based on those from
the Quantronium experiment and they are: $E_{J}=25\times 0.86 k_{b}
K$, $E_{C}=0.68 k_{b} K$, $\text{}E_{\textup{C0}}=0.0037 k_{b} K$,
$E_{\textup{J0}}=18.4 k_{b} K$ and $E_{b}=0.97 E_{\textup{J0}}$;
here the qubit energy $E_{J}$ has been increased by a factor of
25 in order to exaggerate the difference between the trajectories
of the $W_{+}( \gamma ,N,t) $ and $W_{-}( \gamma ,N,t) $ terms. In
these plots with the absence of decoherence the Wigner functions
continuously evolve, moving about their respective trajectories
maintaining their height and shape and hence conserving the equal
probabilities of finding the system in either of the two qubit states
which is consistent with the qubit symmetric superposition state
$(|0\rangle +|1\rangle )/\sqrt{2}$. The superimposed trajectories
are shown from time $t=0$ until $t=2T_{0}$}
\label{XRef-FigureCaption-930213955}
\end{figure}

\begin{figure}[hbtp]
\begin{center}
\includegraphics[scale=0.83]{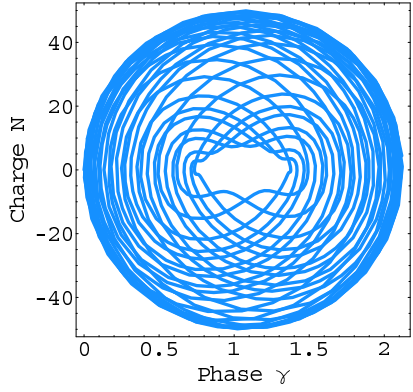}
\end{center}
\caption{Parametric plot of the trajectory of the centre of the
off-diagonal Wigner function term $W_{\times }( \gamma ,N,t) $ for
the parameters used in Fig.~\ref{XRef-FigureCaption-930213955}.
Here the trajectory is shown from $t=0$ to $t=20\times T_{0}$ where
$T_{0}= \pi /(\omega _{+}-\omega _{-})$. From this plot we see the
continual coherence of the Wigner function under the absence of
decoherence, since the off-diagonal $W_{\times }( \gamma ,N,t) $
term continuously evolves in the co-ordinate space without decay. }
\label{XRef-FigureCaption-926175556}
\end{figure}

\section{The Irreversible Current Source - Measurement Induced Decoherence}
\label{XRef-Section-103133835}
\subsection{The Irreversible Current Source Wigner Function Equation}

In the previous section the behaviour of the Quantronium 
system under the application of a bias current, introduced 
as a Hamiltonian term, was examined.  Now we investigate the effect of
an irreversible bias current model by adding the Lindblad derived 
terms that appear in Eq. (\ref{ApproxMasterEquation}) to 
show the relative decoherence in the system's evolution. In 
this case we have the system density matrix $\rho$ defined by the master equation 
\begin{equation}
\overset{.}{\rho }=-\frac{i}{\hbar }[ H_{R}-E_{b}\gamma ,\rho ]
-\frac{|E_{b}|}{2 \hbar }[ \gamma ,\left[ \gamma ,\rho \right] ]\nonumber .
\end{equation}
Using the scaling factors Eq. (\ref{XRef-Equation-910164742})
and Eq. (\ref{XRef-Equation-102193326}) we can write the following 
three master equations to describe the elements of the density 
matrix $\rho $ as
\begin{align}
\overset{.}{\rho }_{+}=&-i\left[ \omega _{+}a_{+}^{\dagger }a_{+}-\frac{E_{b}\Gamma
_{+}}{\hbar }\left( a_{+}^{\dagger }+a_{+}\right) ,\rho _{+}\right] \nonumber \\ &-\frac{|E_{b}|{\Gamma_{+}}^{2}}{2 \hbar }
\left[ a_{+}^{\dagger }+a_{+},\left[ a_{+}^{\dagger}+a_{+},\rho _{+}\right] \right] ,\nonumber
\end{align}
\begin{align}
\overset{.}{\rho }_{-}=&-i\left[ \omega _{-}a_{-}^{\dagger }a_{-}-\frac{E_{b}\Gamma
_{-}}{\hbar }\left( a_{-}^{\dagger }+a_{-}\right) ,\rho _{-}\right]\nonumber \\  &-\frac{|E_{b}|{\Gamma_{-}}^{2}}{2 \hbar }
\left[ a_{-}^{\dagger }+a_{-},\left[ a_{-}^{\dagger}+a_{-},\rho _{-}\right] \right] ,\nonumber
\end{align}
\begin{align}
\overset{.}{\rho }_{\times }=&-i\left[ \omega _{\times }a_{\times }^{\dagger
}a_{\times }-\frac{E_{b}\Gamma _{\times }}{\hbar }\left( a_{\times
}^{\dagger }+a_{\times }\right) ,\rho _{\times }\right]\nonumber \\ &-\frac{|E_{b}|{\Gamma_{\times }}^{2}}{2 \hbar }
\left[ a_{\times }^{\dagger }+a_{\times },\left[
a_{\times }^{\dagger }+a_{\times },\rho _{\times }\right] \right] 
\nonumber\\ &-\frac{i E_{J}
{\Gamma _{\times }}^{2}}{8 \hbar }\left\{ \left( a_{\times }^{\dagger
}+a_{\times }\right) ^{2},\rho _{\times }\right\}.\nonumber
\end{align}
Following the same procedure used in Section \ref{XRef-Section-102193751}
we obtain the three component Wigner function equations. The three uncoupled Wigner function term equations are
\begin{align}
\overset{.}{W}_{+}( \alpha ) =&-i \omega _{+}( \partial _{\alpha
^{*}}\alpha ^{*}-\partial _{\alpha }\alpha ) W_{+}( \alpha )\nonumber \\ 
& -\frac{|E_{b}|{\Gamma _{+}}^{2}}{2 \hbar
}\left( \partial _{\alpha ^{*}}^{2}+\partial _{\alpha }^{2}-2\partial
_{\alpha }\partial _{\alpha ^{*}}\right) W_{+}( \alpha ) \nonumber\\
&-\frac{i
E_{b}\Gamma _{+}}{\hbar }\left( \partial _{\alpha }-\partial _{\alpha
^{*}}\right) W_{+}( \alpha ) ,
\label{XRef-Equation-929212612}
\\
\overset{.}{W}_{-}( \alpha ) =&-i \omega _{-}( \partial _{\alpha
^{*}}\alpha ^{*}-\partial _{\alpha }\alpha ) W_{-}( \alpha )\nonumber \\ &-\frac{|E_{b}|{\Gamma _{-}}^{2}}{2 \hbar
}\left( \partial _{\alpha ^{*}}^{2}+\partial _{\alpha }^{2}-2\partial
_{\alpha }\partial _{\alpha ^{*}}\right) W_{-}( \alpha ) \nonumber \\
& -\frac{i
E_{b}\Gamma _{-}}{\hbar }\left( \partial _{\alpha }-\partial _{\alpha
^{*}}\right) W_{-}( \alpha ) ,
\label{XRef-Equation-929212634} 
\end{align}
\begin{align}
\overset{.}{W}_{\times }( \gamma ,N) =&-\omega _{\times }\left( \lambda
_{\times }\partial _{\gamma }N-\frac{1}{\lambda _{\times }}\partial
_{N}\gamma \right) W_{\times }( \gamma ,N)\nonumber \\ &-\frac{E_{b}}{\hbar }\partial
_{N}W_{\times }( \gamma ,N)+\frac{|E_{b}|}{2 \hbar }\partial _{N}^{2}W_{\times
}( \gamma ,N) \nonumber\\
&-\frac{iE_{J}}{16 \hbar }\left( 4\gamma ^{2}-\partial _{N}^{2}\right)
W_{\times }( \gamma ,N) ,
\label{XRef-Equation-917114031}
\end{align}
where the simplified $\alpha $ notation convention from the previous
section has again been used. By solving these three equations we 
can investigate the evolution of the Quantronium device in the 
presence of the irreversible bias current and in particular the decay 
of the off-diagonal term $W_{\times }$, that projects the qubit into 
one of its eigenstates with probabilities related to the initial state 
of the qubit. 

\subsection{The Irreversible Current Source Wigner Function Solution}
\label{XRef-Subsection-10219413}
\begin{figure}[hbtp]
\begin{center}
(a) \includegraphics[scale=0.83]{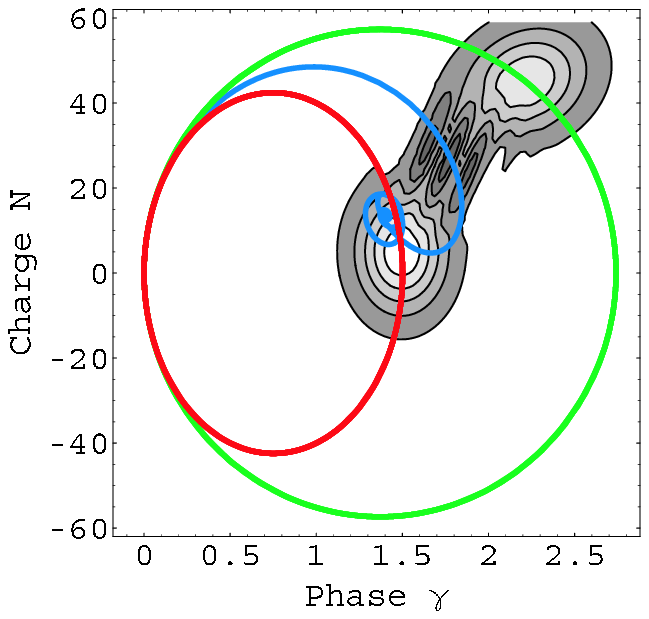}\\
(b) \includegraphics[scale=0.83]{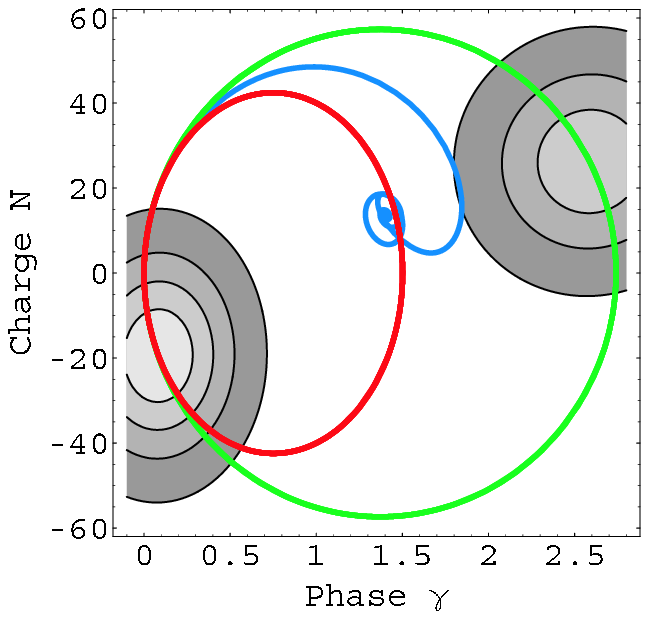}
\end{center}
\caption{Contour plots of the Wigner function $W_{s}( \gamma ,N,t)
$ in the presence of the decoherence from the irreversible bias
current. (a) Contour plot at time $t=T_{0}/4$. (b) Contour plot
at time $t=T_{0}$ where $T_{0}=\pi /(\omega _{+}-\omega _{-})$.
The superimposed trajectories show the time evolution of the centre
of the three Wigner function components $W_{+}( \gamma ,N,t) $,
$W_{-}( \gamma ,N,t) $ and $\operatorname{Re}( W_{\times }( \gamma
,N,t) ) $ shown in red, green and blue respectively. The parameters
used in this plot are identical to Fig.~\ref{XRef-FigureCaption-930213955}
and derived from the Quantronium experiment and again the qubit
energy $E_{J}$ has been increased by a factor of 25 in order to
exaggerate the difference between the trajectories of the $W_{+}(
\gamma ,N,t) $ and $W_{-}( \gamma ,N,t) $ terms. In these plots
we note that the role of the irreversible bias current decoherence
on the Wigner functions. The qubit state which is initially the
symmetric superposition state $(|0\rangle +|1\rangle )/\sqrt{2}$
evolves to a state where the coherence term $W_{\times }( \gamma
,N,t) $ decays to zero and corresponds to a classical equal probability
mixture of the two states $|0\rangle $ and $|1\rangle $. During
the decay of the off-diagonal Wigner function term $W_{\times }(
\gamma ,N,t) $ the trajectory of its centre spirals towards a central
point as the coherence of the qubit state is lost. As the coherence
is lost the diagonal Wigner function terms $W_{+}( \gamma ,N,t)
$ and $W_{-}( \gamma ,N,t) $broaden out and become less localised.
The superimposed trajectories are shown from time $t=0$ until $t=2T_{0}$.}
\label{XRef-FigureCaption-103123754}
\end{figure}

As was the case for the Hamiltonian evolution of the Quantronium
system with the reversible bias current term, which we described
in the previous section, the first two Wigner function equations
(Eq. (\ref{XRef-Equation-929212612}) and Eq. (\ref{XRef-Equation-929212634}))
can be solved using the Wang and Uhlenbeck solution for a linear
Fokker-Plank equation since the equations are in the form of Eq.
(\ref{XRef-Equation-102194049}) where
\begin{gather}
M_{\pm }=\left( \begin{array}{cc}
 i \left( \omega _{\pm }\pm \frac{E_{J} {\Gamma _{\pm }}^{2}}{4 \hbar
}\right)  & i \frac{E_{J} {\Gamma _{\pm }}^{2}}{4 \hbar } \\
 i \frac{E_{J} {\Gamma _{\pm }}^{2}}{4 \hbar } & i \left( \omega _{\pm
}\pm \frac{E_{J} {\Gamma _{\pm }}^{2}}{4 \hbar }\right) 
\end{array}\right) , \nonumber\\ 
N_{\pm }=\frac{1}{\hbar }\left( \begin{array}{cc}
 -|E_{b}|{\Gamma _{\pm }}^{2} & +|E_{b}|{\Gamma _{\pm }}^{2} \\
 +|E_{b}|{\Gamma _{\pm }}^{2} & -|E_{b}|{\Gamma _{\pm }}^{2}
\end{array}\right) , \nonumber \\\nabla _{z}=\binom{\partial _{\widetilde{\alpha
}}}{\partial _{\widetilde{\alpha }^{*}}},z=\binom{\widetilde{\alpha
}}{\widetilde{\alpha }^{*}},\nonumber
\end{gather}
and $\widetilde{\alpha }=\alpha -E_{b}\Gamma_{\pm} /(\hbar
\omega_{\pm} )$. The solution for the diagonal Wigner function terms $W_{+}$
and $W_{-}$ is the again the Gaussian 
\begin{equation}
W_{\pm }( \alpha _{\pm },t) =\frac{1}{2\pi |C_{\pm }|}\exp \left( -\frac{1}{2}{u_{\pm
}}^{T}.C_{\pm }^{-1}.u_{\pm } \right),
\label{XRef-Equation-83016920}
\end{equation}
where 
\begin{equation}
u_{\pm }( \alpha _{\pm },t) =\left(
\begin{array}{c}
	 \alpha _{\pm }-\frac{E_{b}\Gamma_{\pm }}{\hbar  \omega _{\pm }}-
	 e^{-i \omega _{\pm }t}\left( \alpha _{0}-\frac{E_{b}\Gamma_{\pm }}{\hbar  \omega_{\pm }}\right) \\
	 \alpha _{\pm }^{*}-\frac{E_{b}\Gamma_{\pm }}{\hbar  \omega _{\pm }}-
	 e^{+i \omega _{\pm }t}\left( \alpha _{0}^{*}-\frac{E_{b}\Gamma_{\pm }}{\hbar  \omega _{\pm }}\right) \\
\end{array} \right)	\nonumber
\end{equation}
and from the initial condition Eq. (\ref{XRef-Equation-911221614})
we have $\alpha _{0}=0$ and the covariance matrix
\begin{gather}
C_{\pm }=\left( \begin{array}{cc}
 \frac{i |E_{b}|{\Gamma _{\pm }}^{2}}{4 \hbar  \omega _{\pm }}\left(
1- e ^{-2 i \omega _{\pm } t}\right)  & \frac{|E_{b}| {\Gamma _{\pm}}^{2} t }{2 \hbar } \\
 \frac{|E_{b}| {\Gamma _{\pm }}^{2} t}{2\hbar } & \frac{i |E_{b}|{\Gamma_{\pm }}^{2}}{4 \hbar  \omega _{\pm }}\left( e ^{2 i \omega _{\pm
} t}-1\right) 
\end{array}\right) \nonumber\\+\left( \begin{array}{cc}
  e ^{-i \omega _{\pm } t} & 0 \\
 0 &  e ^{+i \omega _{\pm } t}
\end{array}\right) .C_{0}.\left( \begin{array}{cc}
  e ^{-i \omega _{\pm } t} & 0 \\
 0 &  e ^{+i \omega _{\pm } t}
\end{array}\right) 
\label{XRef-Equation-10313159}
\end{gather}
which decays from the initial covariance matrix $C_{0}$
given by Eq. (\ref{XRef-Equation-102194935}). The solutions 
$W_{+}$ and $W_{-}$ (and $W_{\times}$) are shown in Fig.~
\ref{XRef-FigureCaption-103123754}, the parameters used in
this figure are the same as those used in 
Fig.~\ref{XRef-FigureCaption-930213955} and
Fig.~\ref{XRef-FigureCaption-926175556}, where the parameters of the
Quantronium experiment have been used to demonstrate the evolution
of the Wigner function with the exception that the qubit energy
$E_{J}$ has been increased by a factor of 25 to exaggerate the separation
of the states in the presence of the increasing noise characteristics
of the irreversible bias current.

In Fig.~\ref{XRef-FigureCaption-103123754} we see that the trajectories
of the diagonal Wigner terms $W_{+}$ and $W_{-}$ through the $(\gamma
,N)$ co-ordinate space are identical to those found using the reversible
bias current approach of the previous section. With the application
of the bias current $I_{bias}$ the two Gaussians start from the
initial condition where they superimposed on each other at the origin
and then $W_{+}$ and $W_{-}$ separate as they begin to rotate about
the points $E_{b}\Gamma _{+}/\hbar \omega _{+}$ and $E_{b}\Gamma
_{-}/\hbar \omega _{-}$ on the phase axis with a frequency $\omega
_{+}$ and $\omega _{-}$ respectively. However, the shape and hence
the noise characteristics of the $W_{+}$ and $W_{-}$ terms have
changed. The off-diagonal elements $|E_{b}|{\Gamma _{\pm }}^{2}t/2\hbar$ 
in the covariance matrix Eq. (\ref{XRef-Equation-10313159}) mean
that during the evolution of the states, energy from the system
is `leaking' and causing the Gaussians to become broader as they
separate. For long time this means that the states become virtually
indistinguishable in the $(\gamma ,N)$ co-ordinate space. 

For the off-diagonal term $W_{\times }(\gamma ,N ) $ we
solve Eq. (\ref{XRef-Equation-917114031}) with a solution in the 
non-positive definite form Eq. (\ref{XRef-Equation-103122512}). Using this form
of solution we can derive the set of six coupled differential equations
Eq. (\ref{XRef-Equation-103123419}) - Eq. (\ref{XRef-Equation-103123443})
where $G=-E_{b}/\hbar $, $I=-2|E_{b}|/ \hbar $ and $E=E_{J}/4\hbar$. 
We can solve this set of equations numerically using the same
initial conditions used in the previous section, i.e. Eq. (\ref{XRef-Equation-911221614}).
In Fig.~\ref{XRef-FigureCaption-103123754} we see that $W_{\times}$
decays as the system evolves, so that by the time the
diagonal terms $W_{+}$ and $W_{-}$ are the most separated at time
$T_{0}$ it has virtually decayed to zero relative to the diffusing
and larger diagonal Wigner terms. From this numerical solution 
of the off-diagonal Wigner function
term $W_{\times }$ we are now in a position where we can examine
the time it takes for the coherence of the initial qubit symmetric
superposition state to be lost. Also the effect that this description
of the bias current as a Poisson distributed kick process has on
the qubit when white noise in current source is considered.

\section{Decoherence in the Quantronium Experiment}

\subsection{Dephasing Time of Read-Out Current Pulse}

In Section \ref{XRef-Section-102193751} and Section \ref{XRef-Section-103133835}
we have numerically obtained the solution for the off-diagonal Wigner
function term $W_{\times }$ both with and without the presence of
decoherence from the irreversible bias current. These numerical
solutions were obtained in terms of the functions $a( t) $, $b(
t) $, $c( t) $, $d( t) $, $e( t) $ and $f( t) $ that specify the
off-diagonal wigner function in the form of Eq. (\ref{XRef-Equation-103122512}).
From these solutions we can determine the role of the irreversible
bias current on the coherence time of the qubit when it is initially
prepared in a symmetric superposition state. We calculate the coherence time of the qubit
from the length of the Bloch vector ${\mathcal B}( t) $ for the
state of the system defined by
\begin{equation}
{\mathcal B}( t) =\sqrt{\left\langle  \sigma _{x}\right\rangle 
^{2}+\left\langle  \sigma _{y}\right\rangle  ^{2}+\left\langle 
\sigma _{z}\right\rangle  ^{2}}.\nonumber
\end{equation}
For our coupled readout Josephson Junction and Qubit system we 
can write the expectation of some operator $A$ which operates on the Qubit as 
\begin{equation}
\left\langle  A\right\rangle  =\operatorname{Tr}( A \rho ) =\textup{Tr}_{\textup{JJ}}(
\textup{Tr}_{\textup{Qu}}( A \rho ) ).\nonumber
\end{equation}
Since the Wigner quasi-probability distribution function\cite{wigner} allows 
us to compute expectations of operators straightforwardly, that is,
\begin{equation}
\left\langle  A\right\rangle  =\operatorname{Tr}( A \rho ) =\int
_{-\infty }^{+\infty }\int _{-\infty }^{+\infty }A( \gamma ,N) W(
\gamma ,N) d\gamma dN \nonumber
\end{equation}
where $A( \gamma ,N) $ is the Wigner transform of the
operator $A$, we therefore have 
\begin{align}
\left\langle  A\right\rangle  =\int _{-\infty }^{+\infty }\int _{-\infty
}^{+\infty }& d\gamma  dN (W_{+}\left\langle  0\right|A\left|0\right\rangle  +W_{-}\left\langle
1\right|A\left|1\right\rangle  \nonumber \\&+W_{\times }\left\langle  0\left|A\right|1\right\rangle
+{W_{\times }}^{*}\left\langle  1\left|A\right|0\right\rangle )
\label{XRef-Equation-104154238}
\end{align}
for some operator $A$ acting on the qubit. From this expression we can obtain
the expectation values of the Pauli matrices and they are
\begin{gather}
\left\langle  \sigma _{x}\right\rangle  =\int _{-\infty }^{+\infty
}\int _{-\infty }^{+\infty }\operatorname{Re}( W_{\times}) d\gamma dN\nonumber
\\\left\langle  \sigma _{y}\right\rangle  =\int _{-\infty }^{+\infty
}\int _{-\infty }^{+\infty }\operatorname{Im}( W_{\times }) d\gamma dN,\nonumber
\end{gather}
and
\begin{equation}
\left\langle  \sigma _{z}\right\rangle  =\frac{1}{2}\int _{-\infty
}^{+\infty }\int _{-\infty }^{+\infty }\left( W_{+}
-W_{-} \right) d\gamma dN=0.\nonumber
\end{equation}
From these expectation values the length of the Bloch Vector for the qubit can be written as
\begin{equation}
{\mathcal B}( t) =\left|\int _{-\infty }^{+\infty }\int _{-\infty }^{+\infty
}W_{\times }( \gamma ,N,t) d\gamma dN\right|.
\label{XRef-Equation-104154722}
\end{equation}
Since $W_{\times}( \gamma ,N) $ 
is in the form Eq. (\ref{XRef-Equation-103122512})
we can integrate this analytically and then find ${\mathcal
B}( t) $ using the numerical results for the functions $a( t)
$, $b( t) $, $c( t) $, $d( t) $, $e( t) $ and $f( t) $, doing so we have
\begin{align}
{\mathcal B}( t) =&\left|\frac{2 \pi }{\sqrt{4 c( t) e( t) -d( t) ^{2}}}\exp\left[f(t)\right]\right.
\nonumber \\ &\phantom{-}\left.\times\exp\left[ \frac{b( t) ^{2} c( t) -a( t)  b( t)  d( t) +a( t) ^{2} e(
t) }{d( t) ^{2}-4 c( t)  e( t) }\right]\right|.\nonumber
\end{align}

\begin{figure}[hbtp]
\begin{center}
(a)\includegraphics{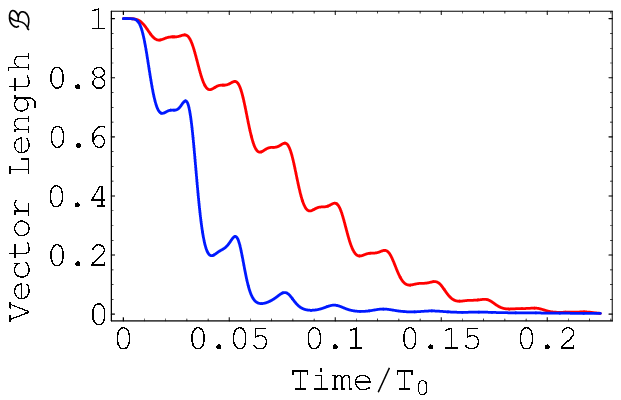}\\
(b)\includegraphics{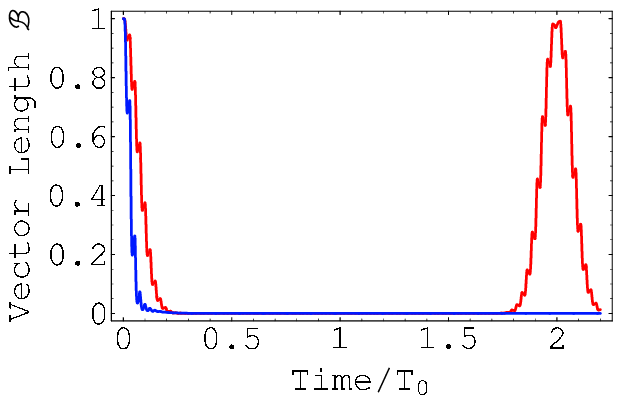}
\end{center}
\caption{Plots of the qubit state Bloch vector length ${\mathcal
B}( t) $ versus time for the decoherence free case (shown in red)
and according to the irreversible bias current model (shown in blue).
The time scale is parameterised in terms of the diagonal Wigner
function term maximum separation time $T_{0}$. Here we note that
in the case of the decoherence free plot the qubit state is a pure
state at time $2n T_{0}$ where $n$ is a positive integer.}
\label{XRef-FigureCaption-10713554}
\end{figure}

In Fig.~\ref{XRef-FigureCaption-10713554} we show the
qubit Bloch vector length as it evolves in time for both the decoherence
free and irreversible bias current solutions. In these plots the
parameters of the Quantronium experiment have been used, and the
graph shows the system evolution from an initial qubit symmetric
superposition state when the manipulation of the qubit has ceased
and the time scale starts at the instant the read-out bias current
pulse is applied. 

The main feature of the decoherence free plots
is the periodic nature of ${\mathcal B}( t) $, in the absence of
decoherence the state of the qubit evolves from an initial pure
state to a mixed state and then back to a pure state when the two
diagonal Wigner function terms $W_{+}$ and $W_{-}$ are superimposed
on each other at time $2T_{0}$. With the application of the irreversible
bias current to the Quantronium experiment we see that the state's
progression to a mixed state is hastened and there is no revival
of the qubit state back to a pure state at time $2T_{0}$. From this
plot we can see that the Bloch vector length is 0.5 at time $0.034T_{0}$, or 0.18 nanoseconds, after
the bias current read-out pulse has been applied to the system.

In the Quantronium experiment the readout pulse lasts for a duration
of the order of 0.1 microseconds, meaning that according to our
irreversible bias current model the state has been dephased on a
time scale that is around a thousand times faster, before any classical
information about the state has been returned to the experimentalist.
The consequences of this is that in the experiment the qubit decoheres 
much faster than the time taken for the measurement.

\subsection{Decoherence Due to Thermal Fluctuations in the Current Source}
\label{XRef-Subsection-108212243}
The irreversible bias current model presented so far not only has
a dephasing effect when the readout current pulse is applied to
the Quantronium circuit but also during the presence of noise in
the current source. Here we look at the decoherence that our model
predicts during the qubit manipulation stage when the mean of the
readout current source is zero but has white noise fluctuations
due to thermal noise in the resistor network which is used in conjunction
with the voltage source in the Quantronium experiment to implement
the readout current pulse\cite{cottetThesis}. In this regime we
consider the full master equation that includes the irreversible
bias current decoherence term
\begin{equation}
\overset{.}{\rho }=-\frac{i}{\hbar }\left[ H,\rho \right] +\frac{i E_{b}}{\hbar
}\left[ \gamma ,\rho \right] -\frac{\left|E_{b}\right|}{2\hbar }\left[ \gamma ,\left[ \gamma
,\rho \right] \right] ,\nonumber
\end{equation}
and we write this in the operator form\cite{stace:062308}
\begin{equation}
\overset{.}{\rho }={\mathcal L}_{H}\left\{ \rho \right\} +\epsilon
_{a}{\mathcal L}_{a}\left\{ \rho \right\} +\epsilon _{b}{\mathcal
L}_{b}\left\{ \rho \right\}
\label{XRef-Equation-823141425}
\end{equation}
where
\begin{gather}
{\mathcal L}_{H}\left\{ \rho \right\} =-\frac{i}{\hbar }[ H,\rho ] ,
\label{XRef-Equation-823143038}
\\\epsilon _{a}{\mathcal L}_{a}\left\{ \rho \right\} =\frac{i E_{b}}{\hbar
}[ \gamma ,\rho ] ,
\end{gather}
and
\begin{equation}
\epsilon _{b}{\mathcal L}_{b}\left\{ \rho \right\} =-\frac{|E_{b}|}{2\hbar
}[ \gamma ,\left[ \gamma ,\rho \right] ] .\nonumber
\end{equation}
Here $\epsilon _{a}$ and $\epsilon _{b}$ are small parameters
compared to the qubit Hamiltonian $H$ and they are determined by
the zero-bias current noise where $\epsilon _{a}=E_{b}$ and $\epsilon
_{b}=|E_{b}|$. We assume $E_{b}$ fluctuates due to thermal 
noise in the external circuit, having zero mean. Therefore $E_{b}$ 
denotes the time fluctuating and zero mean current noise. 
For statistical purposes we treat $\epsilon_{a}$ and 
$\epsilon_{b}$ as independent variables. Since $\langle \epsilon_{b}\rangle = \langle |E_{b}|\rangle \neq 0$, we ignore fluctuations in $| E_{b}|$, whilst $\langle \epsilon_{a}\rangle = \langle E_{b}\rangle =0$
and so we consider the effect of thermal fluctuations in $E_{b}$. 
We write the solution to Eq. (\ref{XRef-Equation-823141425}) as a correction to the exact 
solution $\rho _{0}$ of the noise free master equation
\begin{equation}
\overset{.}{\rho _{0}}=-\frac{i}{\hbar } \left[H,\rho _{0}\right]\nonumber ,
\end{equation}
that is, we can write the solution in the form $\rho =\rho
_{0}+\epsilon _{a}\rho _{a}+\epsilon _{b}\rho _{b}+ O( \epsilon
^{2}) $. That is, $\rho_{a}$ and $\rho_{b}$ represent the change
of the density matrix due to the effect of fluctuations
in the coherent term and for the intrinsic dissipative term 
(that depends on $|E_{b}|$) respectively. Substituting this 
solution $\rho $ into Eq. (\ref{XRef-Equation-823141425})
and expanding the master equation to $O(\epsilon _{a})$ and $O(\epsilon _{b})$ gives
\begin{gather}
\overset{.}{\rho }_{0}={\mathcal L}_{H} \rho _{0},
\label{XRef-Equation-9271336}
\\\overset{.}{\rho }_{a}={\mathcal L}_{H} \rho _{a}+{\mathcal L}_{a}
\rho _{0},
\label{XRef-Equation-9271326}
\end{gather}
\begin{equation}
\overset{.}{\rho }_{b}={\mathcal L}_{H} \rho _{b}+{\mathcal L}_{b}
\rho _{0}.
\label{XRef-Equation-92713223}
\end{equation}
Defining $\widetilde{\rho }_{a}=\rho _{0}+\epsilon_{a} \rho _{a}$ and
$\widetilde{\rho }_{b}=\rho _{0}+\epsilon _{b}\rho _{b}$ we find
\begin{equation}
\overset{.}{\widetilde{\rho }}_{a}=-\frac{i}{\hbar }[ H,\widetilde{\rho
}_{a}] +\frac{i E_{b}}{\hbar }[ \gamma ,\widetilde{\rho
}_{a}] 
\label{XRef-Equation-107165755}
\end{equation}
and
\begin{equation}
\overset{.}{\widetilde{\rho }}_{b}=-\frac{i}{\hbar }\left[ H,\widetilde{\rho
}_{b}\right] -\frac{|E_{b}|}{2\hbar }\left[ \gamma ,\left[ \gamma ,\widetilde{\rho
}_{b}\right] \right] 
\label{XRef-Equation-107172830}
\end{equation}
This allows us to treat the intrinsic dephasing of the current source and 
the dephasing due to thermal fluctuations in the washboard potential independently.
Since $E_{b}$ fluctuates, the second term of 
Eq. (\ref{XRef-Equation-107165755}) results in extra decoherence, on top
of the intrinsic decoherence due to current passing through 
the readout Josephson junction described by
Eq. (\ref{XRef-Equation-107172830}). From these equations we can
estimate the magnitude arising from these different effects.

Assuming the thermal fluctuations of $E_{b}$ are well approximated 
by white noise with zero mean we derive a master equation from 
Eq. (\ref{XRef-Equation-107165755}) which describes the dephasing
effect of a fluctuating current. By integrating Eq.
(\ref{XRef-Equation-107165755}) and substituting into the original master equation
we have
\begin{align}
\overset{.}{\widetilde{\rho }}_{a}=&-\frac{i}{\hbar }\left[ H,\widetilde{\rho
}_{a}\right]\nonumber \\ &+\frac{i}{\hbar }\left[ E_{b}( t) \gamma ,-\frac{i}{\hbar }\int
_{0}^{t}ds \left[H-E_{b}( s) \gamma ,\widetilde{\rho }_{a}(
s) \right] \right] .
\label{XRef-Equation-819155057}
\end{align}
Taking the ensemble average of this equation and using
$\langle E_{b}( t) \rangle =0$ and
$\langle E_{b}( t) E_{b}( s) \rho ( s) \rangle =\langle E_{b}( t)
E_{b}( s) \rangle \langle \rho ( s) \rangle $ (since $\langle \rho
( t) \rangle $ is independent of future noise fluctuations) then
\begin{align}
\langle  \overset{.}{\widetilde{\rho }}_{a}\rangle 
=&-\frac{i}{\hbar }\left[ H,\left\langle  \widetilde{\rho }_{a}\right\rangle
\right] \nonumber \\ &-\frac{1}{\hbar ^{2}}\int _{0}^{t}ds\left\langle  E_{b}( t) E_{b}(
s) \right\rangle  \left[ \gamma , \left[ \gamma ,\left\langle  \widetilde{\rho
}_{a}( s) \right\rangle  \right] \right] .\nonumber
\end{align}
The noise correlation function satisfies
\begin{align}
\left\langle  E_{b}( t) E_{b}( s) \right\rangle  =&\left( \frac{\Phi
_{0}}{2 \pi }\right) ^{2}\left\langle  I_{b i a s}( t) I_{b i a
s}( s) \right\rangle   \nonumber \\=&\left( \frac{\Phi _{0}}{2 \pi }\right) ^{2}{S_{I}}^{2}(
0)\delta ( t-s) ,\nonumber
\end{align}
where by definition
\begin{equation}
{S_{I}}^{2}( \omega )  =\frac{1}{2 \pi }\int _{-\infty }^{+\infty }e^{i
\omega  t}\left\langle  I_{b i a s}( t) I_{b i a s}( 0) \right\rangle
dt \nonumber
\end{equation}
is the noise spectrum of the current fluctuations due
to thermal noise. Thus, in the presence of noise, the master equation
is of the form of Eq. (\ref{XRef-Equation-107172830}). That is, it
can be written as
\begin{equation}
\langle  \overset{.}{\widetilde{\rho }}_{a}\rangle 
=-\frac{i}{\hbar }\left[ H,\left\langle  \widetilde{\rho }_{a}\right\rangle
\right] - \left( \frac{S_{I}( 0)  \Phi _{0}}{2 \pi  \hbar }\right)
^{2} \left[\gamma , \left[ \gamma ,\left\langle  \widetilde{\rho }_{a}\right\rangle
\right] \right] .\nonumber
\end{equation}
Now for the Quantronium circuit with a current source output
resistance $R_{1}$ at temperature $T$ and an effective 
input resistance of $R_{2}$, the current noise spectrum in terms 
of the thermal voltage noise spectrum\cite{nyquist} 
$S_{V}( \omega ) $ is
\begin{equation}
S_{I}( 0) =\frac{S_{V}( 0) }{R_{2}}=\frac{\sqrt{4 R_{1} k_{b} T}
}{R_{2}}.
\label{XRef-Equation-10720542}
\end{equation}

The ensemble average of Eq. (\ref{XRef-Equation-107172830})
is
\begin{equation}
\langle  \overset{.}{\widetilde{\rho }}_{b}\rangle 
=-\frac{i}{\hbar }\left[ H,\left\langle  \widetilde{\rho }_{b}\right\rangle
\right] -\frac{\left\langle  \left|I_{b i a s}\right|\right\rangle  \Phi _{0}}{4
\pi  \hbar }\left[ \gamma ,\left[ \gamma ,\left\langle  \widetilde{\rho
}_{b}\right\rangle  \right] \right] ,\nonumber
\end{equation}
which establishes that the two kinds of dephasing have the same form.
The rates due to the intrinsic dephasing of the current source 
and noise in the washboard potential are given by 
$\gamma_{deph}=\left\langle \left|I_{b i a s}\right|\right\rangle \Phi_{0}/(4\pi\hbar)$ and
$\gamma_{noise}=(S_{I}(0)\Phi_{0}/(2\pi\hbar))^{2}$ respectively. We have so far assumed
that $E_{b}$ is $\delta$-correlated. Under this white noise assumption 
$\langle |E_{b}|\rangle$ is singular so instead we estimate it from 
$I_{\textup{RMS}}$.  The thermal noise in the current source is a result of the random
scattering of electrons in the output resistance and this produces
a statistical distribution of phonons in the phonon modes of the
resistor. In the thermal state each phonon mode has a Gaussian distribution
for the resultant voltage and hence current fluctuations. Summing
over the phonon mode distributions in the resistor we obtain
the total current fluctuations which is also Gaussian distributed.
For Gaussian processes
\begin{equation}
\left\langle  \left|I_{b i a s}\right|\right\rangle  =\sqrt{\frac{2}{\pi }}I_{\textup{RMS}}.\nonumber
\end{equation}
and therefore we have
\begin{equation}
I_{\textup{RMS}}=\sqrt{ \left\langle  {I_{bias}}
^{2}\right\rangle  }=\frac{\sqrt{4 R_{1} k_{b} T B}}{R_{2}}\nonumber
\end{equation}
where $B$ is the bandwidth of the Quantronium circuit.

We can now compare the sizes of the two dephasing terms by referring
to the details of the Quantronium experiment\cite{cottetThesis}.
Analysing the readout circuit we can see that the thermal noise
is produced by a $10\operatorname{k\Omega }$ resistor in series
and a $50\Omega $ resistor in parallel with an ideal voltage source
at the temperature of the helium bath. Also the thermal noise from
these two resistors contribute to the fluctuating current that flows
into the Quantronium Circuit via a $3.5 \operatorname{k\Omega }$
input resistance which has a $200 \operatorname{MHz}$ bandwidth.
From these parameters we are able to determine that 
\begin{equation}
\gamma_{noise}=\left( \frac{S_{I}( 0)  \Phi _{0}}{2 \pi  \hbar }\right) ^{2}=9.68
\operatorname{GHz} \nonumber
\end{equation}
and
\begin{equation}
\gamma_{deph}= \frac{\left\langle  |I_{b i a s}|\right\rangle  \Phi _{0}}{4 \pi
\hbar }=\frac{I_{\textup{RMS}}\Phi _{0}}{\sqrt{8\pi ^{3}} \hbar
}=555 \operatorname{MHz},\nonumber
\end{equation}
meaning that the dephasing rate intrinsic to the irreversible bias current
is about 20 times slower than the rate due to fluctuations in 
the titled washboard potential. From the relative scale of these
two terms we can see that the dephasing during the qubit operation
will be dominated by the fluctuations in the washboard potential, rather
than the intrinsic irreversible bias current induced dephasing. However, 
we note that the introduction 
of the irreversible current source into the modelling process has still provided a 
dephasing effect of considerable 
size relative to the effect of thermal noise in the current source in this case, 
and therefore may be important for the consideration of other similar current 
biased superconducting circuit experimental models.

\section{Discussion and Conclusion}

In this paper we have analysed the bias current readout process
of the superconducting qubit structures such as the Quantronium
(and by analogy those qubits whose quantum state is measured by a DC-SQUID like the
persistent current qubit). By introducing an irreversible bias current term
through Lindblad operators that describe the addition and subtraction
of electrons across the readout Josephson junction, at a rate given by the
bias current, we are able to obtain a master equation that can be
approximated to first order by the Hamiltonian washboard potential
model - a model that is used throughout the superconducting quantum
device literature. Therefore this master equation incorporates an additional
term to the washboard potential terms that dictate the decoherence
of the qubit through its coupling to the readout Josephson junction.

The decoherence is a result of the bias current `counting' the number of electrons that pass through the measurement Josephson junction. We propose that such an effect is intrinsic to the application of the bias current to the system and has a cumulative effect of decohering the system as electrons pass through the readout Josephson junction.  By approximating
the Hamiltonian terms by a harmonic oscillator coupled to a qubit
in the symmetric superposition state were are able to analyse the
measurement induced decoherence before a tunnelling process
out of the washboard potential occurs and produces a measurable
voltage for the experimentalist. Looking at this model in terms of
the Quantronium experiment we have been able to construct the Wigner
function for the Josephson junction and analyse the dephasing effect
upon the application of the external bias current. 

By analysing the Quantronium system we have found that the effect
of describing the readout bias current in terms of the Lindblad
operators is to produce a qubit dephasing time of 0.2 nanoseconds
after the bias current has been applied. In the Quantronium experiment
the bias current pulse was applied for a duration of the order of
0.1 microseconds, meaning that the state of the qubit has been reduced
to a mixed state before the the tunnelling event from the washboard
potential is observed. Our model changes the understanding of the
measurement process of the Quantronium qubit and it means that the
point of measurement is not the tunnelling event out of the washboard
potential but instead arises as a consequence of coupling a current
biased Josephson junction to the qubit state. Additionally this model
does not produce extra sensitivity to noise in the current source since by adding
thermal noise to the irreversible bias current model we showed that
thermal noise in the washboard potential produces the dominant dephasing
effect by an order of magnitude.

Experimental validation of our model could be predicted by using
small current pulses during the Ramsey fringe experiment demonstrated
by Vion \textit{et al.} \cite{Vion} since the role of the irreversible
bias current is to dephase the qubit. Small amplitude and short
duration current pulses could be applied to the Quantronium between
$\pi /2$ pulses of a Ramsey fringe experiment. Our model would predict
that for larger current and longer duration pulses the dephasing
would become larger and hence influence the decay time seen in the
Ramsey fringes. In addition to this the process shown in this paper
of adding the irreversible bias current to the current biased Josephson
junction qubits\cite{{XLiScience301-809, Martinis}} could be employed
to look at the effect of the constant current through the Josephson
junction and its resulting decoherence. In this case due to the
utilisation of excited states of the washboard potential for the
qubit states and readout, an appropriate replacement to the harmonic
oscillator used in this paper would need to be employed.

\section{Acknowledgements}

We acknowledge fruitful discussions with H. S. Goan and C. A. C.
Schelpe. GDH acknowledges the financial support of the Australian
Research Council Special Research Centre for Quantum Computer Technology,
Churchill College and the Cambridge Commonwealth Trust. This work
was supported by the EPSRC and DTI under a Foresight LINK project.

\appendix

\section{Analytical Calculation of the Off-Diagonal Wigner Function
with Reversible Current Source}

Using the commutation relation $[\gamma ,N]=i$ which gives $e^{-i\theta
N}|\gamma \rangle =|\gamma +\theta \rangle $ and the definition
of the Wigner function
\begin{equation}
W( \alpha ,\alpha ^{*}) =\int _{-\infty }^{+\infty }e^{\eta ^{*}
\alpha -\eta  \alpha ^{*}}\mathrm{Tr}\left\{ \rho e^{\eta  a^{\dagger }-\eta
^{*} a}\right\} d^{2}\eta ,\nonumber
\end{equation}
we can write
\begin{equation}
W( \gamma ,N) =\frac{2}{\pi }\int _{-\infty }^{+\infty }e^{i\eta
_{x} N}\left\langle  \gamma -\eta _{x}/2|\rho |\gamma +\eta _{x}/2\right\rangle
d\eta _{x}.
\label{XRef-Equation-9611589}
\end{equation}
Using this form we can calculate the off-diagonal Wigner
function $W_{\times }( \gamma ,N) $ that corresponds to the density
matrix component $\rho _{\times }( t) $ when we decompose the combined
qubit and detector density matrix into the form:
\begin{align}
 \rho ( t) =&\rho _{+}( t) \left|0\right\rangle  \left\langle  0\right|+\rho
_{-}( t) \left|1\right\rangle  \left\langle  1\right| \nonumber \\&+\rho _{\times }( t) \left|0\right\rangle
\left\langle  1\right|+{\rho _{\times }}^{\dagger}( t) \left|1\right\rangle  \left\langle
0\right|. \nonumber
\end{align}
We are able to calculate the off-diagonal Wigner function
$W_{\times }( \gamma ,N) $ in terms of the wavefunctions $|\psi
_{0}( t) \rangle $ and $|\psi _{1}( t) \rangle $ which correspond
to the single mode Gaussian wavefunctions of the Hamiltonian $H_{R}$
in the qubit eigenstates $|0\rangle $ and $|1\rangle $ respectively
since 
\begin{equation}
\rho _{\times }( t) =|\psi _{0}(t)\rangle \langle \psi _{1}(
t) |.\nonumber
\end{equation} 

The wavefunctions $|\psi _{0}( t) \rangle $ and $|\psi _{1}(
t) \rangle $ evolve according to the Hamiltonians $H_{R+}$ and $H_{R-}$
respectively where
\begin{equation}
H_{R\pm }=\hbar  \omega  a^{\dagger }a\pm \hbar  \chi  \left( a+a^{\dagger
}\right) ^{2}+\hbar  \epsilon  \left( a+a^{\dagger }\right) .\nonumber
\end{equation}
The $\gamma $ space wave function for the most general
single mode Gaussian pure state is
\begin{align}
\left\langle  \gamma \big{|}\psi ( t) \right\rangle  =&\left( 2\pi \left\langle
\left( \Delta \gamma \right) ^{2}\right\rangle  \right) ^{-1}\exp
( i\delta _{\gamma }/2) \exp ( -iN_{0}\gamma _{0}/2) \nonumber \\ &\times\exp ( iN_{0}\gamma
) \exp ( -\sigma \left( \gamma -\gamma _{0}\right) ^{2}/2) 
\label{XRef-Equation-96115718}
\end{align}
where
\begin{gather}
\gamma _{0}=\left\langle  \gamma \right\rangle  ,\nonumber
\\N_{0}=\left\langle  N\right\rangle  ,\nonumber
\\\sigma =\sigma _{1}+i\sigma _{2},\nonumber
\\\sigma _{1}=\frac{1}{2\left\langle  \left( \Delta \gamma \right)
^{2}\right\rangle  }\nonumber
\end{gather}
and
\begin{equation}
	\sigma_{2}= -\frac{\langle \Delta \gamma \Delta N\rangle_{\mathrm{sym}}}{2 \langle (\Delta \gamma)^{2}\rangle}=- \frac{\langle N\gamma\rangle +\langle\gamma N\rangle -2 \langle\gamma\rangle\langle N\rangle}{2 \langle (\Delta 	\gamma)^{2}\rangle}.\nonumber
\end{equation}
The phase angle $\delta _{\gamma }$ is set to zero. By 
using the single mode Gaussian form Eq. (\ref{XRef-Equation-96115718})
and the Wigner function definition Eq. (\ref{XRef-Equation-9611589})
we can calculate the off-diagonal Wigner function 
\begin{gather}
W_{\times }( \gamma ,N) =\frac{2}{\pi }\int _{-\infty }^{+\infty
}e^{i\eta _{x}N}\left\langle  \gamma -\frac{\eta _{x}}{2}\right|\rho _{\times
}\left|\gamma +\frac{\eta _{x}}{2}\right\rangle  d \eta _{x}
\nonumber\\ =\frac{2}{\pi
}\int _{-\infty }^{+\infty }e^{i\eta _{x}N}\left\langle  \gamma
-\frac{\eta _{x}}{2} \Big{|} \psi _{0}\left( t\right) \right\rangle  \left\langle
\psi _{1}( t) \Big{|}\gamma +\frac{\eta _{x}}{2}\right\rangle  d \eta
_{x}. \nonumber
\end{gather}
This integral is in the form
\begin{equation}
	\frac{C}{\sqrt{2 \pi}}\int_{-\infty}^{+\infty} \mathrm{d} x ~ {e}^{\Theta x}{e}^{-\frac{1}{2} \Delta x^{2}} = \frac{1}{\sqrt{\Delta}}\exp{\frac{\Theta^{2}}{2 \Delta}} \nonumber
\end{equation}
where
\begin{gather}
	\Theta = iN -\frac{i}{2}(\langle N \rangle_{\scriptscriptstyle +}+\langle N \rangle_{\scriptscriptstyle -})\nonumber \\								+\frac{1}{2}\sigma_{\scriptscriptstyle +}(\gamma - \langle\gamma\rangle_{\scriptscriptstyle +}) - \frac{1}{2} 		
	{\sigma_{\scriptscriptstyle -}^{*}}(\gamma - \langle \gamma\rangle_{\scriptscriptstyle -}),\nonumber
\end{gather}
\begin{equation}
	\Delta = \frac{1}{4} \sigma_{\scriptscriptstyle +} +\frac{1}{4}{\sigma_{\scriptscriptstyle -}^{*}}\nonumber
\end{equation}
and
\begin{gather}
	C =  \frac{2}{\pi\sqrt[4]{\langle (\Delta \gamma)^{2}\rangle_{\scriptscriptstyle +}\langle (\Delta \gamma)^{2}\rangle_{\scriptscriptstyle -}}}\nonumber\\ 		\times\exp{\left( \frac{i}{2}(\langle N\rangle_{\scriptscriptstyle +} \langle\gamma\rangle_{\scriptscriptstyle +}-\langle N\rangle_{\scriptscriptstyle -} 		\langle\gamma\rangle_{\scriptscriptstyle -}) +i\langle N \rangle_{\scriptscriptstyle +} \gamma- i \langle N\rangle_{\scriptscriptstyle -} 					\gamma\right)}\nonumber\\
	\times\exp{ \left(-\frac{1}{2} \sigma_{\scriptscriptstyle +} (\gamma - \langle\gamma\rangle_{\scriptscriptstyle +})^{2}
	-\frac{1}{2}{\sigma_{\scriptscriptstyle  -}}^{*}(\gamma -\langle\gamma\rangle_{\scriptscriptstyle -})^{2} \right)} .\nonumber
\end{gather}
Here we have used the notation $\langle \gamma \rangle
_{\scriptscriptstyle \pm }$, $\langle N\rangle _{\scriptscriptstyle \pm }$, $\langle (\Delta \gamma )^{2}\rangle
_{\scriptscriptstyle \pm }$ and $\sigma _{\scriptscriptstyle \pm }$ to distinguish the mean and noise
parameters of the single mode Gaussian states $|\psi _{0}( t) \rangle
$ and $|\psi _{1}( t) \rangle $ respectively. 

In order to fully specify the off-diagonal Wigner function we need to calculate the
quantities $\langle \gamma \rangle $, $\langle N\rangle $, $\sigma
$ and $\delta _{x}$ for both the states $|\psi _{0}( t) \rangle
$ and $|\psi _{1}( t) \rangle $ where
\begin{equation}
	\langle\gamma\rangle = \sqrt{\frac{\lambda}{2}} (\langle a\rangle + \langle a^{\dagger}\rangle) \nonumber
\end{equation}
\begin{equation}
	\langle N \rangle = 	-\frac{i}{\sqrt{2 \lambda}} (\langle a\rangle - \langle a^{\dagger}\rangle)\nonumber
\end{equation} 
\begin{align}
	\langle (\Delta\gamma)^{2} \rangle =& \frac{\lambda}{2} (\langle aa\rangle +\langle a^{\dagger}a^{\dagger}\rangle +\langle aa^{\dagger} \rangle +\langle a^{\dagger} a\rangle \nonumber \\ &- 2\langle a\rangle\langle a^{\dagger}\rangle - \langle a\rangle^{2} -\langle a^{\dagger}\rangle^{2})\nonumber
\end{align}
and 
\begin{align}
	\langle \Delta \gamma \Delta N \rangle_{\mathrm{sym}} =&  \langle N\gamma\rangle +\langle\gamma N\rangle -2 \langle\gamma\rangle\langle 			N\rangle  \nonumber \\ =& \frac{i}{2} (\langle a^{\dagger} a^{\dagger}\rangle - \langle aa\rangle + \langle a\rangle\langle a\rangle - \langle 				a^{\dagger}\rangle\langle 	a^{\dagger}\rangle).\nonumber
\end{align}
To calculate these quantities we find the two sets of
equations that solve for $\langle aa\rangle$, $\langle a^{\dagger}a^{\dagger}\rangle$, 
$\langle a^{\dagger}a\rangle$, $\langle aa^{\dagger}\rangle$, 
$\langle a\rangle$ and $\langle a^{\dagger}\rangle$ for the 
two qubit-eigenstate Hamiltonians for the system qubit and 
detector $H_{R\pm }$ in the qubit states $|0\rangle $ and $|1\rangle $ 
respectively. In order to find these we construct the set of six 
Heisenberg equations of motion for each
Hamiltonian using the relation
\begin{equation}
	d A/dt=-i[ A,H_{R\pm }] /\hbar 
\end{equation}
and solve them simultaneously. From our Hamiltonians $H_{R+}$
and $H_{R-}$ respectively we find the set of two coupled differential
equations
\begin{align}
	{{d}\langle a\rangle}/{{d}t}=&-i\omega \langle a\rangle\mp 2i\chi (\langle a^{\dagger}\rangle+ \langle a\rangle) - i\epsilon \nonumber\\\nonumber
   	{{d}\langle a^{\dagger}\rangle}/{\mathrm{d}t}=&+i\omega \langle a^{\dagger}\rangle\pm 2i\chi (\langle a^{\dagger}\rangle+\langle a\rangle) 		+i\epsilon,
\end{align}
which we solve using the initial conditions $\langle a(
0) \rangle =0$ and $\langle a^{\dagger }( 0) \rangle =0$. The four
remaining, coupled equations of motion for $H_{R+}$ and $H_{R-}$
are
\begin{align}
   	{{d}\langle aa\rangle}/{{d}t}=& -2i\left(\omega \pm 2\chi\right)\langle aa\rangle
	\nonumber \\&\nonumber
	 \mp 2i\chi (\langle a^{\dagger}a\rangle +\langle a a^{\dagger}\rangle)-2i\epsilon\langle a\rangle \\
  	{{d}\langle a^{\dagger}a^{\dagger}\rangle}/{{d}t}=& +2i\left(\omega \pm 2\chi\right)\langle a^{\dagger}a^{\dagger}\rangle
	\nonumber \\&\nonumber
	\pm 2i\chi (\langle a^{\dagger} a\rangle +\langle a a^{\dagger}\rangle)+2i\epsilon\langle a^{\dagger}\rangle \\\nonumber
	 {{d}\langle a^{\dagger}a\rangle}/{{d}t}=& \mp 2i\chi (\langle a^{\dagger} a^{\dagger}\rangle -\langle aa\rangle) 
	-i\epsilon(\langle a^{\dagger}\rangle-\langle a\rangle) \\\nonumber
  	{{d}\langle aa^{\dagger}\rangle}/{{d}t}=& \mp 2i\chi (\langle a^{\dagger} a^{\dagger}\rangle -\langle a a\rangle) 
	-i\epsilon(\langle a^{\dagger}\rangle-\langle a\rangle) ,\nonumber
\end{align}

Using the solutions $\langle a\rangle _{t}$ and $\langle
a^{\dagger }\rangle _{t}$ we write the four coupled differential
equations in matrix form $\nabla _{t}=A.x+v$, where $x=(aa,a^{\dagger
}a^{\dagger },a^{\dagger }a,aa^{\dagger },a^{\dagger },a)^{T}$,
$\nabla _{t}$ contains the time derivatives of the components of
$x$, and $v$ contains the terms containing $\langle a\rangle _{t}$
and $\langle a^{\dagger }\rangle _{t}$. We solve this system of
equations by diagonalising the matrix $A$ by forming the matrix
$D$ containing its eigenvectors corresponding to the eigenvalues
\begin{equation}
\left\{ 0,0,+2i\sqrt{\omega ^{2}\pm 4\omega \chi },-2i\sqrt{\omega
^{2}\pm 4\omega \chi }\right\} .\nonumber
\end{equation}
Once in the diagonal form $D^{-1}.\nabla _{t}=D^{-1}A.D.D^{-1}.x+D^{-1}.v$
we can solve the four uncoupled differential equations and then
transform back the solution to the original basis. Using the initial
conditions $\langle aa\rangle _{0}=0,\langle a^{\dagger
}a^{\dagger }\rangle _{0}=0,\langle a^{\dagger }a\rangle _{0}=0,\langle
aa^{\dagger }\rangle _{0}=0$ and $\langle a^{\dagger }a\rangle _{0}=1$
we have the solution for $\gamma _{0}$, $N_{0}$ and $\sigma $:
\begin{equation}
	\langle \gamma \rangle_{\pm} = \frac{\sqrt{2\lambda} \epsilon (\cos{(t\sqrt{\omega^{2}\pm 4\omega\chi})}-1)}{\omega \pm 4 \chi}\nonumber
\end{equation}
\begin{equation}
	\langle N \rangle_{\pm} = -\frac{\sqrt{2}\epsilon \sin{(t\sqrt{\omega^{2}\pm 4 \omega \chi})}}{\sqrt{\lambda (\omega^{2} \pm 4 \omega\chi)}} 			\nonumber
\end{equation}
\begin{equation}
	\sigma_{\pm} = \frac{\omega^{2} \pm 2 \chi (2\omega + i \sqrt{\omega^{2}\pm 4 \omega \chi} \sin{(2t \sqrt{\omega^{2}\pm 4 \omega 					\chi}))}}{\lambda \omega (\omega \pm 2 \chi (1+ \cos{(2t \sqrt{\omega^{2}\pm 4 \omega \chi}))})} \nonumber
\end{equation}
\begin{equation}
	\langle (\Delta\gamma)^{2}\rangle_{\pm}=\frac{\lambda (\omega \pm 2 \chi (1+ \cos{(2t \sqrt{\omega^{2}\pm 4 \omega \chi}))})}{2 ( \omega \pm 4		\chi)}\nonumber
\end{equation}
In this calculation we have set the phase angle $\delta
_{\gamma }$ to zero for both the $|\psi _{0}( t) \rangle $ and $|\psi
_{1}( t) \rangle $ states. Now that we have fully specified the
mean and noise parameters for the two single mode Gaussian sates
$|\psi _{0}( t) \rangle $ and $|\psi _{1}( t) \rangle $ we can write
this solution in the form $W_{\times }( \gamma ,N) =\exp ( a( t)
\gamma +b( t) N+c( t) \gamma ^{2}+d( t) \gamma N+e( t) N^{2}+f(
t) ) $ where
\begin{gather}
	a(t)=	\frac{(i\langle N\rangle_{\scriptscriptstyle +}+i\langle N\rangle_{\scriptscriptstyle -}+\sigma_{\scriptscriptstyle +}
	\langle\gamma\rangle_{\scriptscriptstyle +}- \sigma_{\scriptscriptstyle -}^{*} \langle\gamma\rangle_{\scriptscriptstyle -}) 							(\sigma_{\scriptscriptstyle -}^{*}-\sigma_{\scriptscriptstyle +})}{\sigma_{\scriptscriptstyle +}+{\sigma_{\scriptscriptstyle -}^{*}}}
	\nonumber \\+i\langle N\rangle_{\scriptscriptstyle +}-i\langle N\rangle_{\scriptscriptstyle -}
	+ \sigma_{\scriptscriptstyle +}\langle\gamma\rangle_{\scriptscriptstyle +} 
	+{\sigma_{\scriptscriptstyle -}^{*}} \langle\gamma\rangle_{\scriptscriptstyle -}\nonumber
\end{gather}
\begin{equation}
	b(t)= \frac{2 (\langle N\rangle_{\scriptscriptstyle +}+\langle N \rangle_{\scriptscriptstyle -})-2i\sigma_{\scriptscriptstyle +} 							\langle\gamma\rangle_{\scriptscriptstyle +} + 2i{\sigma_{\scriptscriptstyle -}^{*}} 	
	\langle\gamma\rangle_{\scriptscriptstyle -} }{\sigma_{\scriptscriptstyle +}+{\sigma_{\scriptscriptstyle -}^{*}}}\nonumber
\end{equation}
\begin{equation}
	c(t)=-\frac{1}{2}\sigma_{\scriptscriptstyle +}-\frac{1}{2}{\sigma_{\scriptscriptstyle -}^{*}}+\frac{(\sigma_{\scriptscriptstyle +}-
	{\sigma_{\scriptscriptstyle -}^{*}})^{2}}{2 (\sigma_{\scriptscriptstyle +}+{\sigma_{\scriptscriptstyle -}^{*}})}\nonumber
\end{equation}
\begin{equation}
	d(t)=\frac{2i {\sigma_{\scriptscriptstyle -}^{*}}-2i\sigma_{\scriptscriptstyle +}}{\sigma_{\scriptscriptstyle +}+{\sigma_{\scriptscriptstyle -}^{*}}}\nonumber
\end{equation}
\begin{equation}
	e(t)=-\frac{2}{\sigma_{\scriptscriptstyle +}+{\sigma_{\scriptscriptstyle -}^{*}}}\nonumber
\end{equation}
and
\begin{gather}
	f(t)=\ln{\left(\frac{4}{\pi\sqrt{\sigma_{\scriptscriptstyle+}+{\sigma_{\scriptscriptstyle-}^{*}}}
	\sqrt[4]{\langle (\Delta \gamma)^{2}\rangle_{\scriptscriptstyle+}\langle (\Delta \gamma)^{2}\rangle_{\scriptscriptstyle-}}}\right)}
	\nonumber \\+\frac{i}{2}(\langle N\rangle_{\scriptscriptstyle+}\langle\gamma\rangle_{\scriptscriptstyle+}-\langle N\rangle_{\scriptscriptstyle-} 			\langle\gamma\rangle_{\scriptscriptstyle-})
	\nonumber \\ - \frac{1}{2}\sigma_{\scriptscriptstyle+}{\langle\gamma\rangle_{\scriptscriptstyle+}}^{2}-\frac{1}{2}{\sigma_{\scriptscriptstyle-}^{*}}
	{\langle\gamma\rangle_{\scriptscriptstyle-}}^{2}
	\nonumber \\ +\frac{(i\langle N\rangle_{\scriptscriptstyle+}+i\langle N\rangle_{\scriptscriptstyle-}
	+\sigma_{\scriptscriptstyle+}\langle\gamma\rangle_{\scriptscriptstyle+}
	-\sigma_{\scriptscriptstyle-}^{*} \langle\gamma\rangle_{\scriptscriptstyle-})^{2}} 
	{2(\sigma_{\scriptscriptstyle+}+{\sigma_{\scriptscriptstyle-}^{*}})} \nonumber
\end{gather}

\section{Analytical Calculation of the Off-Diagonal Wigner Function
with Irreversible Current Source}

For the off-diagonal term $W_{\times }( \gamma ,N ) $ of the Wigner
function including measurement induced decoherence we solve (Eq. (\ref{XRef-Equation-917114031}))
with a solution in the non-positive definite form
\begin{align}
W_{\times }( \gamma ,N,t) =\exp ( &a( t) \gamma +b( t) N+c( t) \gamma
^{2}\nonumber \\ &+d( t) \gamma N+e( t) N^{2}+f( t) ) . \nonumber
\end{align}
Using this form of solution we can derive the set of six
coupled differential equations: 
\begin{align}
\overset{.}{a}( t) &=\frac{\omega }{\lambda _{\times }}b( t) +G
d( t) +\frac{\left( -I+i E\right) }{2}b( t) d( t) , \nonumber
\\
\overset{.}{b}( t) &=-\lambda _{\times }\omega  a( t) +2G e( t)
+\left( -I+i E\right) e( t) b( t) , \nonumber
\\
\overset{.}{c}( t) &=\frac{\omega }{\lambda _{\times }}d( t) -i
E+\frac{\left( -I+i E\right) }{4}d( t) ^{2}, \nonumber
\\
\overset{.}{d}( t) &=-2\lambda _{\times }\omega  c( t) +\frac{2\omega
}{\lambda _{\times }}e( t) +\left( -I+i E\right) e( t) d( t) , \nonumber
\\
\overset{.}{e}( t) &=-\lambda _{\times }\omega  d( t) +\left( -I+i
E\right) e( t) ^{2}, \nonumber
\\
\overset{.}{f}( t) &=G b( t) +\frac{\left( -I+i E\right) }{4}\left(
2e( t) +b( t) ^{2}\right) , \nonumber
\end{align}
where $G=-E_{b} /\hbar $, $I=-2|E_{b}|/\hbar
$ and $E=E_{J}/4\hbar $. This system of equations is solved
by first considering the three coupled equations for $\overset{.}{c}(
t) $, $\overset{.}{d}( t) $ and $\overset{.}{e}( t) $ whereby using
the transformation of variables 
\begin{gather}
z = \frac{d + i ( \frac{c}{\lambda_{\times}}- \lambda_{\times} e)}{4i},\nonumber
\\ \bar{z} = \frac{d - i ( \frac{c}{\lambda_{\times}}- \lambda_{\times} e)}{4i},\nonumber
\end{gather}
and
\begin{equation}
 u = \frac{ (\frac{c}{\lambda_{\times}}+ \lambda_{\times} e)}{4}\nonumber
\end{equation}
we have 
\begin{gather}
\overset{.}{z}=2i\omega _{\times }z-\left( -I+i E\right) \left(
z-u\right) ^{2}/2-2iE,\nonumber
\\\overset{.}{\overset{\_}{z}}=-2i\omega _{\times }\overset{\_}{z}+\left(
-I+i E\right) \left( \overset{\_}{z}+u\right) ^{2}/2+2iE,\nonumber
\\\overset{.}{u}=-\left( -I+i E\right) \left( z-u\right) \left(
\overset{\_}{z}+u\right) /2-2iE.\nonumber
\end{gather}
Using the relation $dZ/dt=-(-I+iE)Z(
z-\overset{\_}{z}-2u) $, where $Z=z\overset{\_}{z}+u^{2}-iE/(-I+iE)$,
we use a second transformation of variables
\begin{gather}
U=u/Z,\nonumber
\\A=z/Z\nonumber
\end{gather}
and
\begin{equation}
\overset{\_}{A}=\overset{\_}{z}/Z\nonumber
\end{equation}
so that 
\begin{gather}
\overset{.}{U}=-\frac{\left( -I+i E\right) }{2}\frac{\left( z\overset{\_}{z}+u^{2}+4iE/\left(
-I+i E\right) \right) }{\left( z\overset{\_}{z}+u^{2}-4iE/\left(
-I+i E\right) \right) },\nonumber
\\\overset{.}{A}=2i\omega _{\times }A-\frac{\left( -I+i E\right)
}{2}\frac{\left( z\overset{\_}{z}+u^{2}+4iE/\left( -I+i E\right)
\right) }{\left( z\overset{\_}{z}+u^{2}-4iE/\left( -I+i E\right)
\right) },\nonumber
\\\overset{.}{\overset{\_}{A}}=-2i\omega _{\times }A+\frac{\left(
-I+i E\right) }{2}\frac{\left( z\overset{\_}{z}+u^{2}+4iE/\left(
-I+i E\right) \right) }{\left( z\overset{\_}{z}+u^{2}-4iE/\left(
-I+i E\right) \right) }.\nonumber
\end{gather}
From these equations we construct the differential equation
\begin{equation}
\frac{d^{4}P}{d t^{4}}+4 \omega _{\times }^{2} \frac{d^{2} P}{d
t^{2}}-16 \omega _{\times }^{2} i E( -I+i E) P=0\nonumber
\end{equation}
where
\begin{equation}
P=\frac{\left( z \overset{\_}{z}+u^{2}+4i E/\left( -I+i E\right)\nonumber
\right) }{\left( z \overset{\_}{z}+u^{2}-4i E/\left( -I+i E\right)
\right) }
\end{equation}
The solution $P$ is the sum of exponentials 
\begin{equation}
P=C_{1}e^{\lambda _{1}t}+C_{2}e^{-\lambda _{1}t}+C_{3}e^{\lambda
_{2}t}+C_{4}e^{-\lambda _{2}t}\nonumber
\end{equation}
where 
\begin{equation}
\lambda _{1,2}=\sqrt{-2\omega _{\times }^{2}\pm 2\omega _{\times
}\sqrt{\omega _{\times }^{2}- 4 {E^{2}}-4i E I}}\nonumber
\end{equation}
and 
\begin{gather}
C_{1,2} = \frac{ {\lambda_2}^2 ( 2I -iE \pm E(E+4iI)/(2\lambda_1)}
{4I({\lambda_2}^2-{\lambda_1}^2)},\nonumber
\\C_{3,4} = -\frac{ {\lambda_1}^2 ( 2I -iE \pm E(E+4iI)/(2\lambda_2)}
{4I({\lambda_2}^2-{\lambda_1}^2)}.\nonumber
\end{gather}
From $P$ we have the solutions for $c( t) $, $d( t) $
and $e( t) $ 
\begin{gather}
c(t) = \frac{(2w^2 \frac{dP}{dt} + \frac{d^2P}{dt^2})}
{4\lambda_{\times} w^2 (-I+iE)(1-P)}\nonumber
\\ d(t) = - \frac{1}{2 w (-I+iE)(1-P)}\frac{d^2P}{dt^2}\nonumber
\\ e(t) = \frac{\lambda_{\times}}{2 (-I+iE)(1-P)}\frac{dP}{dt}.\nonumber
\end{gather}
If $I=0$, $\lambda_{1,2}$ are pure imaginary and so
$P $ simply oscillates resulting in oscillatory solutions for
$c(t)$, $d(t)$, $e(t)$. The remaining coeffcients $a(t)$ and 
$b(t)$ are coupled together, satisfying a forced, parametrically 
excited second order ordinary differential equation. To see this let
\begin{align}
a(t)&= \frac{i}{2 ( 1-P(t))\sqrt{2 \lambda
_{\times}}} \frac{d ( y(t) \sqrt{1-P(t)})}{dt} \nonumber\\
 b(t) &= \sqrt{\frac{\lambda
_{\times}}{2}}\left(-\frac{i y(t)}{2 \sqrt{1-P(t)}} + 
\frac{4 G}{(-I + iE)}\right). \nonumber
\end{align}
Then $y(t)$ satifies the following equation.
\begin{equation}
	\frac{d^2 y}{dt^2} + 
\left( 1-\frac12 \left(\frac{1}{(1-P(t))} \frac{dP(t)}{dt}\right)^2 \right)y(t)
= -\frac{4i G \sqrt{1-P(t)}}{-I+iE} \nonumber
\end{equation}
Since initially both $a(0)=0$ and $b(0)=0$ their subsequent solution
is proportional to $G$ and
\begin{equation}
	\frac{b^2(t) c(t) - a(t) b(t) d(t)+ a^2(t) e(t)}{4 c(t) e(t) - d^2(t)} \nonumber
\end{equation}
is proportional to $G^2$. Initially this is zero and for small times is 
quadratic in time. The solution to $f(t)$ is found through integration.

\end{document}